\documentclass[fleqn,usenatbib]{mnras}

\usepackage{newtxtext,newtxmath}
\usepackage{graphicx}
\usepackage{subcaption}
\usepackage[T1]{fontenc}

\DeclareRobustCommand{\VAN}[3]{#2}
\let\VANthebibliography\thebibliography
\def\thebibliography{\DeclareRobustCommand{\VAN}[3]{##3}\VANthebibliography}

\usepackage{graphicx}	% Including figure files
\usepackage{amsmath}	% Advanced maths commands

\defcitealias{FallowsSanders2024}{FS24}

\title[ ]{Towards Galactic Archaeology with Inferred Ages of Giant Stars From \textit{Gaia} Spectra }

\author[]{Aisha S. Almannaei,$^{1}$\thanks{E-mail: aisha.almannaei.18@ucl.ac.uk (KTS)}
Daisuke Kawata,$^{1,2}$
Ioana Ciuc{\u{a}},$^{3,4,5}$
Connor Fallows,$^{6}$
Jason~L.~Sanders,$^{6}$ %\orcidlink{0000-0003-4593-6788}
\newauthor
George Seabroke,$^{1}$
Andrea Miglio$^{7,8}$
\\
$^{1}$Mullard Space Science Laboratory, University College London, Holmbury St Mary, Dorking, Surrey RH5 6NT, UK\\
$^{2}$National Astronomical Observatory of Japan, 2-21-1 Osawa, Mitaka, Tokyo 181-8588, Japan\\
$^{3}$Kavli Institute for Particle Astrophysics \& Cosmology (KIPAC), Stanford University, Stanford, CA 94305, USA\\
$^{4}$Research School of Astronomy and Astrophysics, Australian National University, Canberra, ACT 2611, Australia\\
$^{5}$School of Computing, Australian National University, Canberra, ACT 2601, Australia\\
$^{6}$Department of Physics and Astronomy, University College London, Gower Street, London WC1E 6BT, UK\\
$^{7}$Dipartimento di Fisica e Astronomia, Universita´ degli Studi di Bologna, Via Gobetti 93/2, I-40129 Bologna, Italy\\
$^{8}$INAF – Osservatorio di Astrofisica e Scienza dello Spazio, Via P. Gobetti 93/3, I-40129 Bologna, Italy
}

\date{Accepted XXX. Received YYY; in original form ZZZ}

\pubyear{2015}

\begin{document}
\label{firstpage}
\pagerange{\pageref{firstpage}--\pageref{lastpage}}
\maketitle

\begin{abstract}
In the era of \textit{Gaia}, the accurate determination of stellar ages is transforming Galactic archaeology. We demonstrate the feasibility of inferring stellar ages from \textit{Gaia}'s RVS spectra and the BP/RP (XP) spectrophotometric data, specifically for red giant branch and high-mass red clump stars. We successfully train two machine learning models, dubbed \texttt{SIDRA}: Stellar age Inference Derived from \textit{Gaia} spectRA to predict the age. The \texttt{SIDRA-RVS} model uses the RVS spectra and \texttt{SIDRA-XP} the stellar parameters obtained from the XP spectra. Both models use \texttt{BINGO}, an APOGEE-derived stellar age as the training data. \texttt{SIDRA-RVS} estimates ages of stars whose age is around $\tau_\mathrm{BINGO}=10$~Gyr with a standard deviation of residuals of $\sim$ 0.12 dex in the unseen test dataset, while \texttt{SIDRA-XP} achieves higher precision with residuals $\sim$ 0.064 dex for stars around $\tau_\mathrm{BINGO}=10$~ Gyr. Since \texttt{SIDRA-XP} outperforms \texttt{SIDRA-RVS}, we apply \texttt{SIDRA-XP} to analyse the ages for 2,218,154 stars. This allowed us to map the chronological and chemical properties of Galactic disc stars, reproducing the known distinct features such as the Gaia-Sausage-Enceladus merger and a potential gas-rich interaction event linked to the first infall of the Sagittarius dwarf galaxy. This study demonstrates that machine learning techniques applied to Gaia's spectra can provide valuable individual age information, particularly for giant stars, thereby enhancing our understanding of the Milky Way's formation and evolution. 

\end{abstract}

\begin{keywords}
Galaxy: stellar content -- Galaxy: abundances– asteroseismology -- methods: data analysis-statistical

\end{keywords}

%%%%%%%%%%%%%%%%%%%%%%%%%%%%%%%%%%%%%%%%%%%%%%%%%%

%%%%%%%%%%%%%%%%% BODY OF PAPER %%%%%%%%%%%%%%%%%%

\section{Introduction}
\label{sec:intro}
With the advent of large spectroscopic surveys, such as the Apache Point Observatory Galactic Evolution Experiment \citep[APOGEE,][]{Majewski2017}, Galactic Archaeology with HERMES \citep[GALAH,][]{DeSilva2015}, Large Sky Area Multi-Object Fiber Spectroscopic Telescope \citep[LAMOST,][]{Cui2012} and the precise astrometry from the \textit{Gaia} mission \citep{GaiaCollab2016, GaiaCollabDR32023}, precise determination of stellar chemical abundances, motions and estimated ages—known as chrono-chemodynamical information—is revolutionising Galactic archaeology. It marks an initial stride in empirically comprehending the formation and evolution of our Galaxy that have led to its current structure.

Chrono-chemodynamical information will provide us answers to questions on the chronological order of the formation of the thick and thin Galactic disc and how to define them \citep{Fuhrmann1998, Fuhrmann2011, Chiappini2015, Ciuca2021, Anders2023}. However, age is a stellar quantity that cannot be directly measured, and one must rely on other physical characteristics of the star that are correlated with age to derive it \cite[see][for a review]{Soderblom2010}. Consequently, age determination is almost always model-dependent, highlighting the need for robust methods.

One way to derive stellar ages from spectroscopic surveys, particularly those with FGK spectral types, involves identifying their position on the Kiel diagram and contrasting that position with stellar evolution models.While this approach provides accurate age estimations based on precise measurements of effective temperature, $T_\mathrm{eff}$, surface gravity, log \textit{g}, and iron abundance ratio, [Fe/H], in regions of the Kiel diagram where isochrones with different ages are distinct \citep[e.g.][]{Bensby2011}, notably at the main-sequence turn-off and the subgiant branch, its reliability is compromised when determining the ages of red giants. This is because on the red giant branch, isochrones of various ages are less distinguishable in the Kiel diagram and to measure the age, more precise stellar parameters are required.

However, obtaining age estimates of brighter giants is also valuable for exploring the Milky Way's formation history and structure. Their substantial luminosity enables observation over extensive distances. This importance stems from the fact that giants in both old and young populations exhibit similar luminosities and colours, leading to a more uniform age selection function compared to turn-off stars. 

An alternative method (but also model-dependent) for determining the ages of stars, including red giants, is through asteroseismology \citep{ChaplinMiglio2013}. Asteroseismology entails directly measuring the seismic frequencies of a star to derive accurate measurements of its internal properties and structure, including radius and mass \cite[see][for a review]{Aerts2021}. Since the mass of the star determines its main sequence lifetime, the mass of a low-mass red giant can effectively reveal its age precisely \citep[e.g., see][]{Miglio2012}. 
% Nevertheless, it requires a significant amount of precise time-series photometry and computational resources, making it feasible only for a limited number of stars at a time. %Instances of missions specifically suitable to conducting asteroseismology, including red giant stars, are
Asteroseismic measurements of the mass and ages of the stars, including giants, are conducted with \textit{Kepler} \citep{Koch2010, Gilliland2010, Bedding2010, Ceillier2017, Yu2018},  Convection, Rotation, and Planetary Transits \citep[CoRoT,][]{Auvergne2009, Miglio2009, Mosser2010, Kallinger2010, Valentini2016}, K2 \citep{Howell2014, Zinn2022} and the Transiting Exoplanet Survey Satellite \cite[TESS,][]{Ricker2014, Mackereth2021, Hon2021}. 
However, asteroseismology requires a significant amount of precise time-series photometry and computational resources, making it feasible only for a limited number of stars at a time.

\citet{MassersonGilmore2015} demonstrated that photospheric carbon and nitrogen abundances (specifically [C/N]) are a good indicator of the mass of giant stars, thereby providing the possibility to estimate their age from stellar spectra. This finding enables the use of extensive spectroscopic survey data, either directly from the spectra or through derived spectroscopic-parameters, to predict stellar ages. However, this method requires calibration with precisely measured ages or masses by asteroseismology. Building on this, \citet{Martig2016} showed that it is possible to infer ages from the well-calibrated APOGEE DR12 spectroscopic derived parameters. They demonstrated that the masses of red-giant stars can be precisely predicted based on their [C/N] abundances, along with their spectroscopic stellar labels, including $T_\mathrm{eff}$, log \textit{g} and [Fe/H] \citep[see also][]{Mackereth2017, Ciuca2021, Anders2023}.  Similarly, \citet{Ness2016} found that CNO molecular lines, specifically CN and CO, within the APOGEE survey also encompass mass-related and, consequently, age-related information and used the spectra directly to predict ages. Extending this approach to other surveys, \citet{Ho2017} illustrated that the CH and CN characteristics in the blue region of the LAMOST spectra are indicative of mass, and hence age \citep[see also][]{He2022}. 

These studies used machine learning techniques to model complex non-linear relationships between input features, such as spectral data or derived spectroscopic parameters, and output results, like stellar age. The machine learning model is first trained on a smaller set of stars with precisely measured ages determined through asteroseismology. Once trained, the model is then used to estimate the ages of a larger population of giant stars for which only spectroscopic data are available.

Presently, \textit{Gaia}'s third data release \citep[DR3,][]{GaiaCollabDR32023} is producing mean Radial Velocity Spectrometer (RVS) spectra for 1 million well-behaved objects, with a significant portion expected to be red giants. These  are all-sky high resolution spectra (approximately R = 11,500), enabling a comprehensive study of Milky Way archaeology, structure and evolution \citep{GaiaCollab2023}. Chemical abundances and atmospheric stellar parameters have been estimated from the RVS spectra by the \textit{Gaia} collaboration \citep{RecioBlanco2023}.

Additionally, low-resolution spectrophotometric measurements, known as \textit{Gaia} BP/RP spectra (hereafter XP spectra), are available for over about 220 million objects as part of the third data release of \textit{Gaia}~DR3.
% mission \citep[DR3,][]{GaiaCollabDR32023}.
These spectra cover the optical wavelength range from 330 nm to 1050 nm with a wavelength-dependent resolution between 30 and 100 \citep{Carrasco2021, DeAngeli2023,Montegriffo2023GaiaCollab}. Despite their low resolution, these spectra provide sufficient information to accurately estimate fundamental stellar parameters, such as effective temperature, surface gravity and metallicity \citep{Liu2012, Witten2022, Andrae2023, Martin2024}, with the added potential to derive more detailed abundance measurements, including elements like carbon and alpha elements \citep{Gavel2021, Witten2022, Lucey2023, SandersMatsunaga2023, Li2024}. Further \citet{Guiglion2024} used machine learning models to derive the stellar parameters more precisely, using the combined information of the RVS spectra, photometry, parallax and the XP data. 

Building on previous studies that extracted detailed abundance information from XP spectra \citep[e.g.][]{Witten2022, Li2024}, \citet[][hereafter FS24]{FallowsSanders2024} developed a feed-forward neural network to predict stellar parameters and abundances using XP spectra, along with photometry from \textit{Gaia}, the Two Micron All Sky Survey \citep[2MASS,][]{2MASSSkrutskie2006} and \textit{Wide-field Infrared Survey Explorer} \citep[unWISE,][]{Schlafly2019}. Their model carefully considers uncertainties in both inputs and outputs, as well as additional model uncertainty. They have published their estimates of  $T_\mathrm{eff}$, log $g$ and various abundance ratios, such as [C/Fe], [Fe/H], [N/Fe] and [$\mathrm{\alpha/M}$].

This paper aims to demonstrate the feasibility of deducing stellar ages from \textit{Gaia}'s RVS spectra and XP stellar parameter obtained from \citetalias{FallowsSanders2024}, particularly for giant stars by training a supervised machine learning model using the training data from the APOGEE giant stars' age data provided by Bayesian INference for Galactic archaeOlogy \citep[\texttt{BINGO},][]{Ciuca2024}. We develop \texttt{SIDRA} (Stellar Age Inference  Derived from \textit{Gaia} SpectrA) to estimate the stellar age of giants from \textit{Gaia} spectra, constructing two models: \texttt{SIDRA-RVS}, which leverages input features from RVS spectra, and \texttt{SIDRA-XP}, which employs input parameters from the XP stellar parameters of \citetalias{FallowsSanders2024}. \textit{Gaia}~DR3 provides the age estimate from the RVS spectra, \texttt{flame\_age\_spec}, and the XP data, \texttt{flame\_age} \citep{Fouesneau2023}. In Appendix \ref{sec:appendixA}, we compare them with the age estimates by \texttt{BINGO}. The stellar age measurements provided in \textit{Gaia}~DR3 show rather limited performance, which demonstrates the necessity of the further improvement. 

The paper is structured as follows: in Section \ref{sec:data}, we present our selection of training \textit{Gaia} DR3 spectral data used in this work; in Section \ref{sec:SIDRA-RVS}, we describe our machine learning model for deriving stellar age using RVS spectra; and in Section \ref{sec:SIDRA-XP}, we detail our machine learning model for deriving stellar age using XP parameters. The outcomes of our models can be found in Section \ref{subsec:RVSresults} for \texttt{SIDRA-RVS} and in Section \ref{subsec:XPresults} for \texttt{SIDRA-XP}. In Section~\ref{sec:chronochemicalmap}, we then apply our machine learning model to all the selected giant stars whose precise XP stellar parameters are available in \citetalias{FallowsSanders2024}, which are not only our APOGEE giant samples, but also the other giants in the \textit{Gaia} data. Section \ref{sec:summary} presents a summary of our work.

%%%%%%%%%%%%%%%%%%%%%%%%%%%%%%%%%%%%%%%%%%%%%%%%%%%%%%%%%%%%%%%%%%%%%%%%
\section{BINGO giants sample }
\label{sec:data}
%%%%%%%%%%%%%%%%%%%%%%%%%%%%%%%%%%%%%%%%%%%%%%%%%%%%%%%%%%%%%%%%%%%%%%%%

\citet{Ciuca2024} estimated the age for red giant branch (RGB) and high-mass red-clump (RC) stars using \texttt{BINGO} as described in \citet{Ciuca2021}. \texttt{BINGO} is a Bayesian Neural Network model designed to map APOGEE stellar parameters, such as $T_\mathrm{eff}$, log \textit{g}, [Fe/H], [Mg/Fe], [C/Fe] and [N/Fe], to stellar age. The model is trained using APOGEE DR17 data in conjunction with \textit{Kepler} asteroseismology data, with stellar parameters from APOGEE DR17 as input features and asteroseismic age data from \citet{Miglio2021} as the output, while accounting for the observational uncertainties during the training process. \citet{Ciuca2024} limited the training data to the RGB stars and high-mass ($>1.8~\mathrm{M}_{\odot}$) RC stars, where asteroseismology can measure the initial mass of the stars confidently, because of their negligible mass loss. 

\citet{Ciuca2024} developed a classification model to identify RC stars with masses exceeding 1.8 $\mathrm{M}_{\odot}$ and RGB stars within the entire APOGEE dataset, restricting their analysis to these specific stellar types. The model for this classification task consists of a three-layer Artificial Neural Network (ANN) developed using \textsc{keras} and \textsc{tensorflow} \citep{abadi2016tensorflow}. It is trained on a subset of RC stars with masses exceeding 1.8 $\mathrm{M}_{\odot}$ and RGB stars, using data from APOGEE DR17 along with asteroseismic mass and evolutionary phase measurements. The trained classifier model was then used to confidently select RC and RGB stars from the APOGEE DR17 data without asteroseismology data. For these selected stars, \texttt{BINGO} was applied to predict their ages. 

We use the APOGEE~DR17 stars whose age is estimated with \texttt{BINGO} for our training and verification data. In this paper, we apply a strict cut on the \texttt{BINGO} age uncertainty, $\sigma_{\log_{10}( \mathrm{\tau_{BINGO}[Gyr]})} < 0.02$ dex, and use only the data selected after this cut. Although the \texttt{BINGO} age is a predicted age from a machine learning model and is subject to both statistical and systematic errors, we consider the \texttt{BINGO} age as the ground truth for our model, making them the reference standard for our age estimations. It is worth noting that the age uncertainties obtained from \texttt{BINGO} imply uncertainties in the knowledge about the model predictions, which may be narrower than the observed uncertainties in the initial asteroseismic age \cite[see][for more details]{Ciuca2021}. It is also worth noting that the ages of some old stars appear to be much older than the age of the Universe. This discrepancy occurs because the asteroseismic age measurements used for the training set in the \texttt{BINGO} model from \citet{Miglio2021} do not incorporate a prior for the maximum possible age, such as the age of the Universe.

\begin{figure*}
    \centering
    \begin{subfigure}[b]{0.495\textwidth}
        \includegraphics[width=\linewidth]{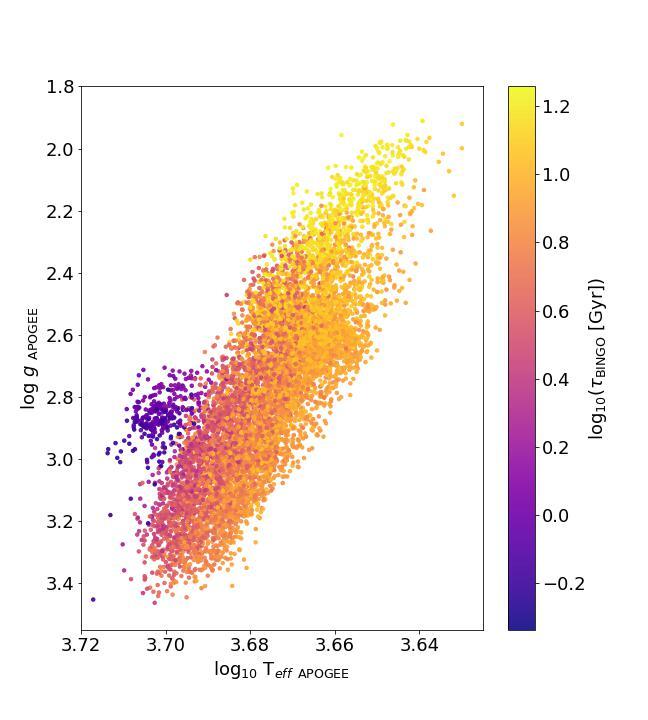}
        % \caption{RVS data cross-matched with \texttt{BINGO}'s APOGEE ID}
      \label{subfig:HRdiag_RVS}
    \end{subfigure}
    \hfill
    \begin{subfigure}[b]{0.495\textwidth}
        \includegraphics[width=\linewidth]{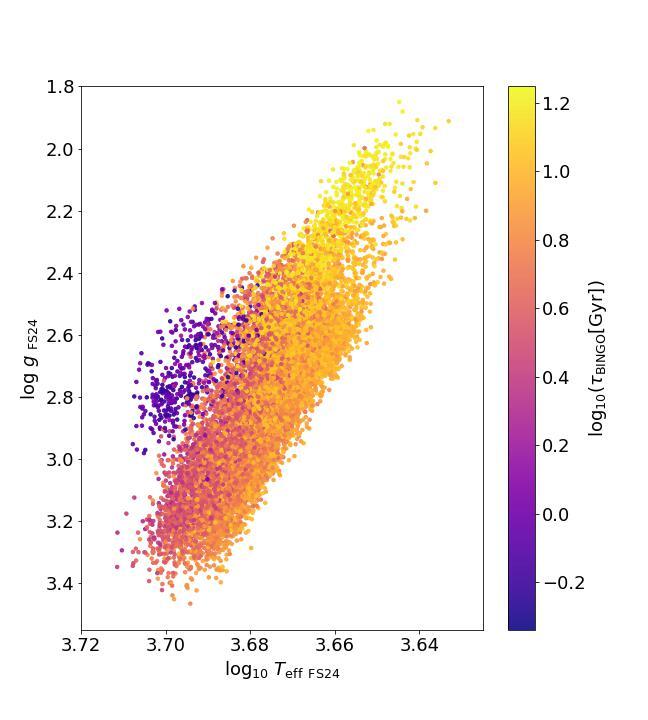}
        % \caption{XP data cross-matched with \texttt{BINGO}'s APOGEE ID}
        \label{subfig:HRdiag_XP}
    \end{subfigure}
    \caption{The Kiel diagram of our cross-matched dataset between the \citet{Ciuca2024} APOGEE data and \textit{Gaia}~DR3 RVS spectra data (left panel) and the \citetalias{FallowsSanders2024} XP stellar parameter data (right panel). The stellar parameters for the RVS data are derived from the APOGEE dataset, while those for the XP data come from \citetalias{FallowsSanders2024}.}
    \label{fig:HRdiag}
\end{figure*}

We cross-match the APOGEE \texttt{BINGO} stars with \textit{Gaia}~DR3 stars with RVS spectra, and obtain 8,859 stars with RVS spectra and high-precision \texttt{BINGO} ages (\texttt{BINGO}-RVS data hereafter). Their distribution in the Kiel diagram is seen in the left panel of Fig. \ref{fig:HRdiag}. The high-mass RC stars (younger stars) are focused mainly around 2.7 $<\log$ $g<$ 3, 3.69 $<\mathrm{log}$ $T_\mathrm{eff}<$ 3.71 with ages between $-0.5<\mathrm{log}_{10}(\tau_\mathrm{BINGO}  \mathrm{[Gyr]})<0.25$. This trend is typical among younger RC stars, because those in the core helium burning phase tend to have higher $\log~g$ compared to the lower mass RC stars. 

We also construct our ground-truth \texttt{BINGO} giant sample to build and evaluate a machine-learning model to estimate the stellar age for the giant stars from \textit{Gaia}'s XP stellar parameters derived by \citetalias{FallowsSanders2024}. The selected dataset from \citetalias{FallowsSanders2024} XP stellar parameter data consists of stars that meet specific accuracy criteria and parameter ranges. We select only the stars whose stellar parameters uncertainties are within their reported median formal uncertainties, i.e. an effective temperature uncertainty, $T_{\mathrm{eff}_\mathrm{unc}}<69.1$ K, surface gravity uncertainty, log $g_{\mathrm{unc}}<0.14$ dex, and metallicity uncertainties, $[\mathrm{Fe}/\mathrm{H}]_{\mathrm{unc}}<0.068$ dex, and $[\mathrm{\alpha}/\mathrm{M}]_{\mathrm{unc}}<0.029$ dex. In cases where duplicate entries exist, the star with the smaller $T_{\mathrm{eff}_\mathrm{unc}}$ is selected. We then cross-match these data with the APOGEE stars with the \texttt{BINGO} ages from \citet{Ciuca2021}, and construct our ground-truth data (a total of 12,500 stars, \texttt{BINGO}-FS24 data hereafter). In this work, we use the \citetalias{FallowsSanders2024} data with only the RVS radial velocity data, as we find the predicted stellar parameters to be more reliable when restricted to this dataset. Additionally, this dataset offers a similar magnitude range to our cross-matched RVS spectra sample, enabling a fair comparison.

Their distribution in the Kiel diagram is seen in the right panel of Fig. \ref{fig:HRdiag}. The figure shows the stellar parameters from \citetalias{FallowsSanders2024}, $\log~g_\mathrm{FS24}$ and $T_\mathrm{eff~FS24}$, colour-coded with stellar ages that are derived from \texttt{BINGO}, $\mathrm{log}_{10}(\tau_\mathrm{BINGO}$  $\mathrm{[Gyr]})$.  It is worth noting that the model used to derive the stellar pameters in  \citetalias{FallowsSanders2024} is trained with the APOGEE data. This explains why the stellar parameters from \citetalias{FallowsSanders2024} (right panel of Fig.~\ref{fig:HRdiag}) reproduce the distribution of the stellar parameters of APOGEE stars (left panel of Fig.~\ref{fig:HRdiag}).

%%%%%%%%%%%%%%%%%%%%%%%%%%%%%%%%%%%%%%%%%%%%%%%%%%%%%%%%%%%%%%%%%%%%%%%%
\section{\texttt{SIDRA} for the RVS spectra: \texttt{SIDRA-RVS}}
\label{sec:SIDRA-RVS}
%%%%%%%%%%%%%%%%%%%%%%%%%%%%%%%%%%%%%%%%%%%%%%%%%%%%%%%%%%%%%%%%%%%%%%%%

The RVS spectra are anticipated to hold valuable age-related information for giant stars, including several nitrogen lines, CN lines, and CO molecular bands \citep{RecioBlanco2023}. Consequently, it may be possible to derive [C/N] from the RVS spectra. If so, a machine learning model could be trained to predict stellar ages using data cross-matched with \texttt{BINGO} age measurements. This section explores the feasibility of training such a machine learning model to infer age directly from the RVS spectra.

\subsection{\texttt{SIDRA-RVS}'s methodology}
\label{subsec:RVSmethods}

A supervised machine-learning regression model is a type of algorithm that is trained on a labeled data set to predict a continuous numerical output or target variable. In a regression task, the aim is to learn the relationship and patterns between the input features and the associated target values, or labels. Once trained properly, the model is able to make precise predictions for new, unseen data. In this case, our input data are the RVS normalised flux obtained from \textit{Gaia} DR3, and our labels are the stellar ages obtained from \citet{Ciuca2024} sample for RGB and high-mass stars as described in Sec. \ref{sec:data}. 

We develop \texttt{SIDRA-RVS} to predict the ages from the \textit{Gaia} spectra of RGB and high-mass RC stars. For the analysis of RVS spectra, we use an eXtreme Gradient Boosting \citep[\texttt{XGBoost},][]{ChenGuestrin2016}, a tree-based machine-learning algorithm that constructs a sequence of decision trees to provide predictions for regression tasks. Each decision tree corrects the errors of the previous ones, contributing to improving the overall model's accuracy. The key abilities and features of this model make it well matched to accurately predict stellar ages. For example, \texttt{XGBoost} exhibits excellent performance in managing extensive data sets, demonstrating notably faster processing speeds than ANNs. It is a flexible algorithm for dealing with missing data instances, making it suitable for the characteristics of our RVS data that sometimes include missing data points in the spectra. RVS spectra are observed within the 846-870 nm filter in the observed wavelength range. Before making the combined spectrum, they are Doppler-shifted to the rest frame. Because 
\textit{Gaia}~DR3 provides the data in a fixed wavelength range in the rest frame, and stars with a large radial velocity have a substantial shift in the wavelength range, one end of the spectrum lead to no data for such stars. There are sometimes a lack of the data in the middle of the spectrum, when all the observed data in that wavelength bin have been corrupted for various reasons, e.g. due to defective pixel. 
Moreover, \texttt{XGBoost} is an ensemble method which combines the predictions of multiple weak regression models or learners known as `trees' to create a more robust and accurate model. This ensemble strategy improves predictive performance and mitigates the risk of overfitting. This was shown by \cite{Borisov2022} who benchmarked various regression algorithms for heterogeneous data sets and found that algorithms based on gradient-boosted tree ensembles perform better than deep-learning models in supervised learning tasks. This model has been utilised for similar previous research, including the work of \citet{Anders2023}, who employed the \texttt{XGBoost} algorithm to assess the age and chemical abundances of APOGEE red-giant stars.

Due to the imbalance of the number of stars between younger and older stars in the dataset for \texttt{SIDRA-RVS}, we augment the training data to make our training data homogeneously distributed as a function of the age. We begin by calculating the kernel density estimation (KDE) distribution of age, $p(\tau)$, for our training dataset. Next, we compute $g_\mathrm{i} = \frac{\max[p(\tau)]}{p(\tau_\mathrm{i})}$ for each star $i$. From this, we generate $N_\mathrm{gen,i} = \text{floor}(g_\mathrm{i})$ additional stars for star $i$. Additionally, a random number, $r_\mathrm{i}$, is drawn between 0 and 1. If $r_\mathrm{i} < g_\mathrm{i} - N_\mathrm{gen,i}$, i.e., smaller than the threshold of $g_\mathrm{i}$, we generate one more star for star $i$. To create additional training star data for star $i$, we draw a Gaussian random value for the star's age, $\tau_j$, based on the logarithmic age, $\log_{10}(\tau_\mathrm{i})$, and its associated uncertainty, $\sigma_{\log_{10}(\tau), \mathrm{i}}$, from the \texttt{BINGO} method, such that 

\begin{equation}
\log_{10}(\tau_j) = \mathcal{N}(\log_{10}(\tau_\mathrm{i}), \sigma_{\log_{10}(\tau), \mathrm{i}}).
\end{equation}

For the spectra generation, we compute the flux, $f_\mathrm{j,\lambda}$, for star $j$, at wavelength, $\lambda$, as 

\begin{equation}
f_\mathrm{j,\lambda} = \mathcal{N}(f_\mathrm{i,\lambda}, \sigma_\mathrm{i,\lambda}),
\end{equation}

where the flux is sampled from a Gaussian distribution using the flux uncertainty, $\sigma_\mathrm{i,\lambda}$, from \textit{Gaia}~DR3 for each flux pixel. We implement \texttt{Optuna} \citep{Akiba2019+Optuna}, a hyperparameter optimisation framework tool for our \texttt{SIDRA-RVS} model to efficiently fine-tune our model and find the best parameters.

%%%%%%%%%%%%%%%%%%%%%%%%%%%%%%%%%%%%%%%%%%%%%%%%%%%%%%%%%%%%%%%%%%%%%%%%
\subsection{Age inference from RVS spectra using \texttt{SIDRA-RVS} }
\label{subsec:RVSresults}
%%%%%%%%%%%%%%%%%%%%%%%%%%%%%%%%%%%%%%%%%%%%%%%%%%%%%%%%%%%%%%%%%%%%%%%%

%%%%%%%%%%%%%%%%%%%%%%%%%%%%%%%%%%%%%%%%%%%%%%%%%%%%%%
%Training results colour-coded by [alpha/M] and [M/H]
%%%%%%%%%%%%%%%%%%%%%%%%%%%%%%%%%%%%%%%%%%%%%%%%%%%%%%

\begin{figure*}
    \centering
    \begin{subfigure}[b]{0.495\textwidth}
        \includegraphics[width=\linewidth]{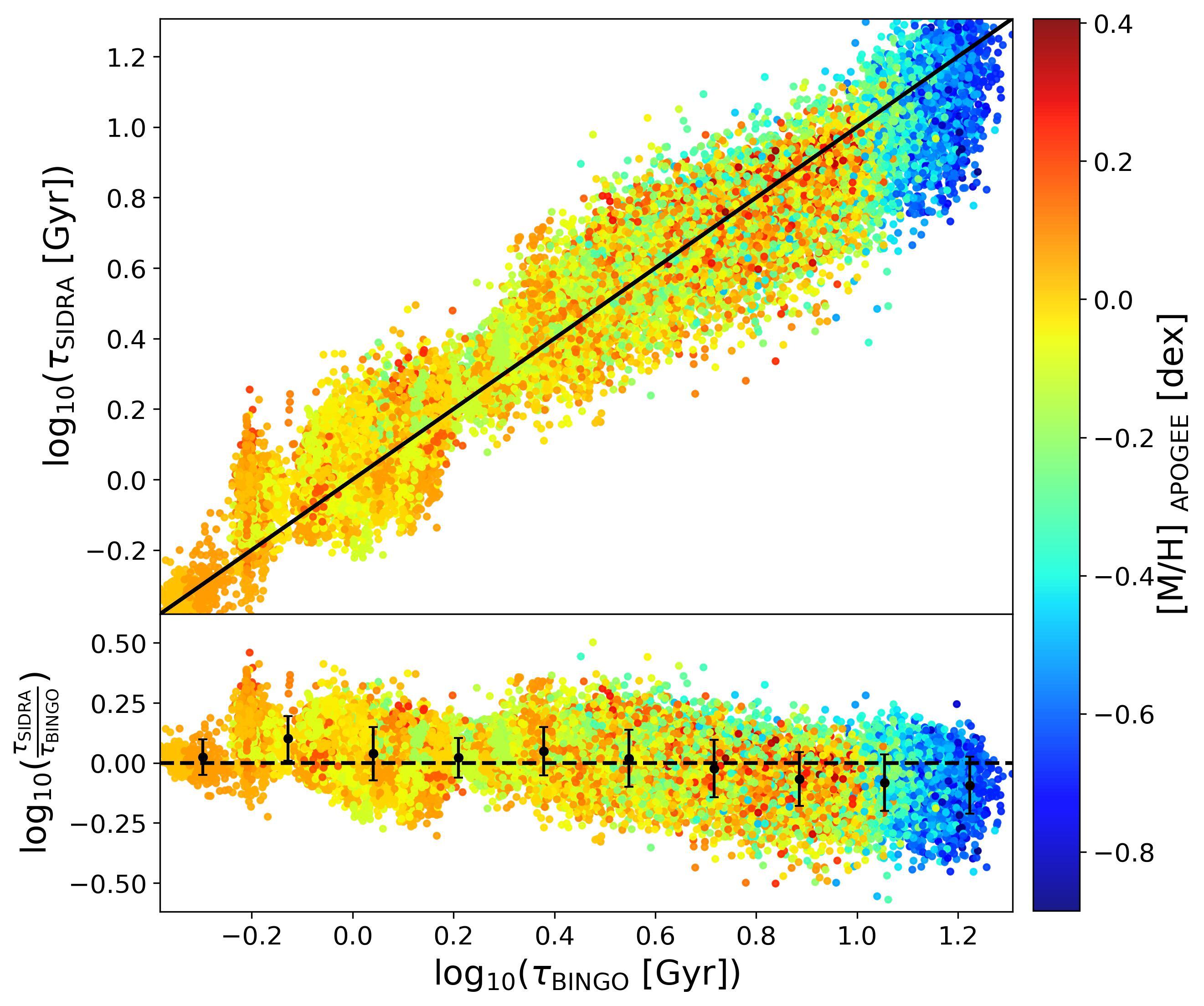}
        \caption{Training set}
        \label{fig:training_results_logage_colored_mh}
    \end{subfigure}
    \hfill
    \begin{subfigure}[b]{0.495\textwidth}
        \includegraphics[width=\linewidth]{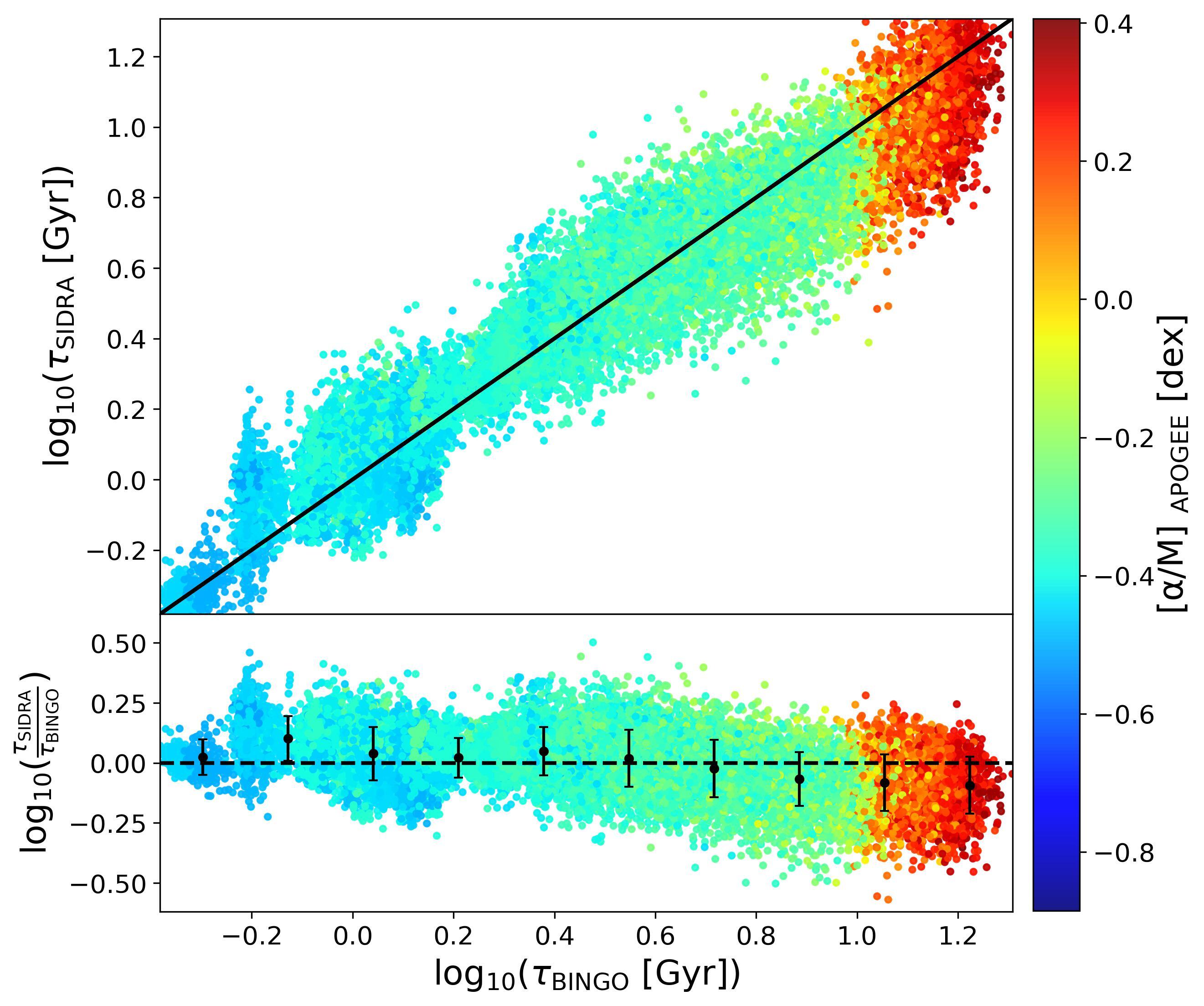}
        \caption{Training set}
        \label{fig:training_results_logage_colored_alpham}
    \end{subfigure}
    \caption{\texttt{SIDRA-RVS} age predictions, $\mathrm{log}_{10}(\tau_\mathrm{SIDRA}$ $\mathrm{[Gyr]})$, versus the target \texttt{BINGO} age estimation in $\mathrm{log}_{10}(\tau_\mathrm{BINGO}$ $\mathrm{[Gyr]})$ for the training set colour-coded by [M/H] (left panel) and [$\mathrm{\alpha/M}$] (right panel) obtained from APOGEE. The upper panels show the predictions versus the target and the black line indicates the identity line. The lower panels represent the residuals between the \texttt{SIDRA-RVS}'s $\mathrm{log}_{10}(\tau_\mathrm{SIDRA}$ $\mathrm{[Gyr]})$ prediction and \texttt{BINGO}'s true $\mathrm{log}_{10}(\tau_\mathrm{BINGO}$ $\mathrm{[Gyr]})$ denoted as $\mathrm{log}_{10}(\frac{\tau_\mathrm{SIDRA}} {\tau_\mathrm{BINGO}})$. The black filled circles and vertical error bars indicate the mean and the standard deviation of the residuals at the different $\mathrm{log}_{10}(\tau_\mathrm{BINGO}$ $\mathrm{[Gyr]})$ bins, respectively.}
    \label{fig:logage_training_results_colorcoded}
\end{figure*}

\begin{figure*}
    \centering
   \begin{subfigure}[b]{0.495\textwidth}
       \includegraphics[width=\linewidth]{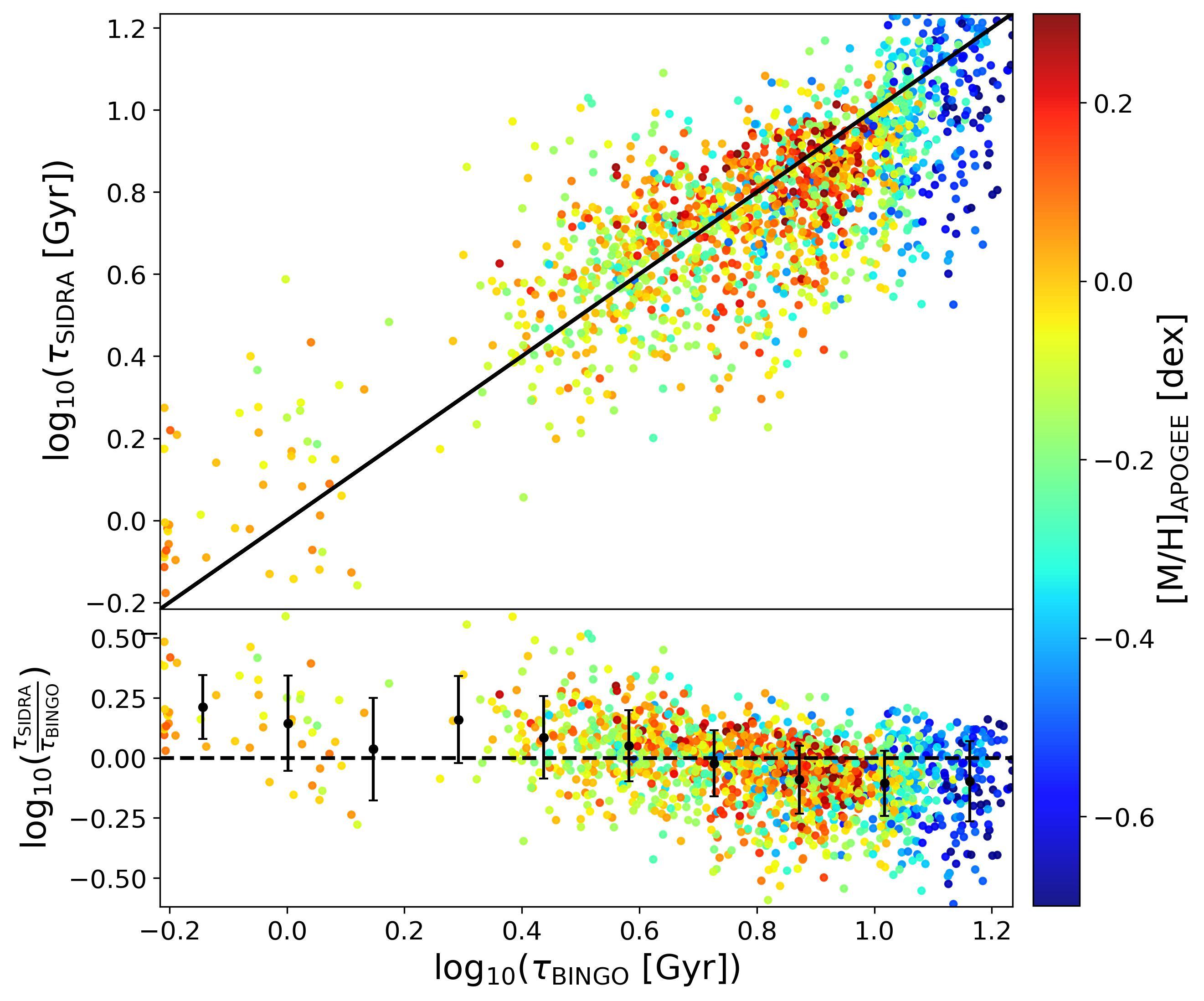}
        \caption{Testing set }
        \label{fig:test_results_logage_colored_MH}
    \end{subfigure}
    \hfill
    \begin{subfigure}[b]{0.495\textwidth}
        \includegraphics[width=\linewidth]{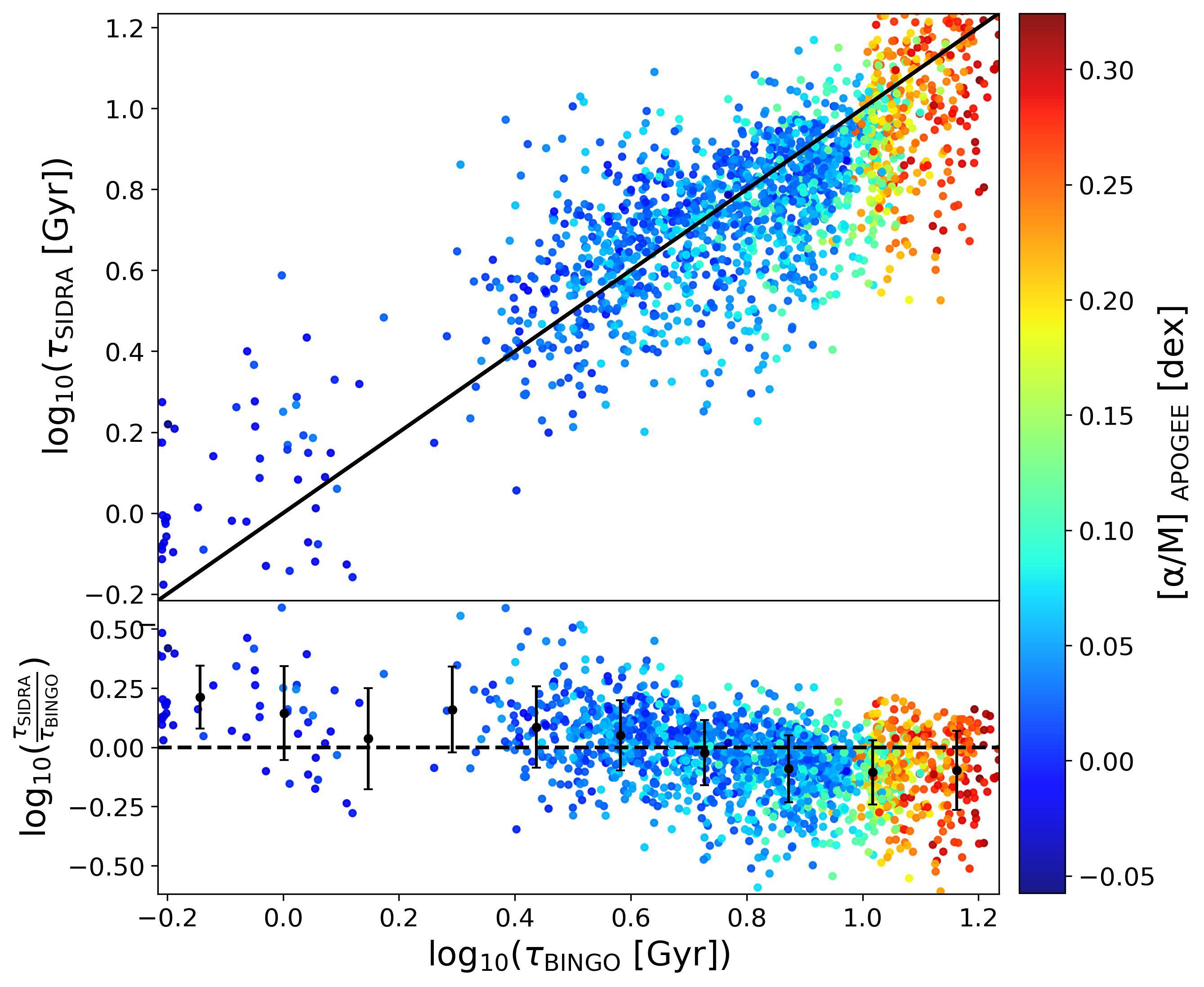}
        \caption{Testing set}
        \label{fig:test_results_logage_colored_alpham}
    \end{subfigure}
    \caption{\texttt{SIDRA-RVS} age predictions, $\mathrm{log}_{10}(\tau_\mathrm{SIDRA}$ $\mathrm{[Gyr]})$ versus the target \texttt{BINGO} age estimation in $\mathrm{log}_{10}(\tau_\mathrm{BINGO}$ $\mathrm{[Gyr]})$ for the testing set colour-coded by [M/H] (left panel) and [$\mathrm{\alpha/M}$] (right panel) obtained from APOGEE. The upper panels show the predictions versus the target and the black line indicates the identity line. The lower panels represent the residuals between the \texttt{SIDRA-RVS}'s $\mathrm{log}_{10}(\tau_\mathrm{SIDRA}$ $\mathrm{[Gyr]})$ prediction and \texttt{BINGO}'s true $\mathrm{log}_{10}(\tau_\mathrm{BINGO}$ $\mathrm{[Gyr]})$ denoted as $\mathrm{log}_{10}(\frac{\tau_\mathrm{SIDRA}} {\tau_\mathrm{BINGO}})$. The black filled circles and vertical error bars indicate the mean and the standard deviation of the residuals at the different $\mathrm{log}_{10}(\tau_\mathrm{BINGO}$ $\mathrm{[Gyr]})$ bins, respectively.}
    \label{fig:logage_results_colorcoded}
\end{figure*}

In this section, we explore the performance of the stellar-age measurements with the RVS spectra from our sample of RGB and high-mass RC stars using \texttt{XGBoost}. Fig. \ref{fig:logage_training_results_colorcoded} illustrates the comparison between the ground-truth \texttt{BINGO} ages, $\mathrm{log}_{10}(\tau _{\mathrm{BINGO}}$ $\mathrm{[Gyr]})$, and the corresponding \texttt{SIDRA-RVS} predicted ages, $\mathrm{log}_{10}(\tau _{\mathrm{SIDRA}}$ $\mathrm{[Gyr]})$, for the stars in the training data, while Fig. \ref{fig:logage_results_colorcoded} shows the same results for the test data. We adopt $80\%$ of our cross-matched \texttt{BINGO}-RVS data for the training data, and the rest of the data are used as the unseen test data after the model is trained. Note that the data augmentation outlined in Sec. \ref{subsec:RVSmethods} is applied exclusively to the training data. As a result, Fig. \ref{fig:logage_training_results_colorcoded} displays a larger and more uniformly distributed dataset with respect to age. In contrast, the test data shown in Fig. \ref{fig:logage_results_colorcoded} illustrates the imbalance in the data. In both the training and testing sets, despite some variations in the predictions, the predictions generally follow the identity line, though a mild systematic trend is present. Using the binned points in the lower panel of Fig. \ref{fig:logage_results_colorcoded}, the residuals show a negative slope of approximately $\text{slope} \approx -0.27 \pm 0.01$, corresponding to a change of roughly 0.1 dex in the residuals from the youngest to the oldest stars, indicating that the \texttt{SIDRA-RVS} ages are slightly compressed relative to the \texttt{BINGO} ages. The bottom panel represents the residuals between \texttt{SIDRA-RVS}'s $\mathrm{log}_{10}(\tau _{\mathrm{SIDRA}}$ $\mathrm{[Gyr]})$ prediction and the ground truth \texttt{BINGO} $\mathrm{log}_{10}(\tau _{\mathrm{BINGO}}$ $\mathrm{[Gyr]})$. In the case of the test data shown in Fig. \ref{fig:logage_results_colorcoded}, the dispersion of residuals is significantly higher for stars with lower ages ($\mathrm{log}_{10}(\tau _{\mathrm{BINGO}}$ $\mathrm{[Gyr]})<0.2$) due to the scarcity of young stars in the test dataset, which leads to a dominance of Poisson noise. In contrast, the residual dispersion for the training data in Fig. \ref{fig:logage_training_results_colorcoded} remains nearly constant across all ages. It is important to note that without the data augmentation applied to the training data, a severe overestimation of the age of young stars is observed in both the training and test data, as there is a lack of young stars ($\mathrm{log}_{10}(\tau _{\mathrm{BINGO}}$ $\mathrm{[Gyr]})<0.2$). While the mean deviation for the test data still indicates an overestimation of age for young stars, the performance is significantly improved compared to a model trained without data augmentation. As a result , the mean of the standard deviation of residual is 0.10 dex for stars ($\mathrm{log}_{10}(\tau _{\mathrm{BINGO}}$ $\mathrm{[Gyr]})>0.2$) in the testing set and there are no strong systematic offsets. The standard deviation of the residuals is around 0.09 dex for the training data and 0.12 dex for the test data around $\mathrm{log}_{10}(\tau_\mathrm{BINGO}$ $\mathrm{[Gyr]})= 1$. This suggests that the model performs similarly on the test set, although a modest increase in scatter remains.

As mentioned in Sec. \ref{sec:data}, our original \texttt{BINGO} data are selected with the quality cut of their statistical uncertainty for $\mathrm{log}(\tau _{\mathrm{BINGO}}$ $\mathrm{[Gyr]}$ being less than 0.02, i.e. $\sigma_{\log( \mathrm{\tau_{BINGO}[Gyr]})} < 0.02$ dex. Our predictions show a larger dispersion compared to the original data. However, this is not surprising, considering the much lower spectral resolution and shorter wavelength coverage of the \textit{Gaia} RVS data than the APOGEE data.

%%%%%%%%%%%%%%%%%%%%%%%%%%%%%%%%%%%%%%%%%%%%%%%%
%Test results colour-coded by [alpha/M] and [M/H]
%%%%%%%%%%%%%%%%%%%%%%%%%%%%%%%%%%%%%%%%%%%%%%%%

The colours of the dots in Figs. \ref{fig:logage_training_results_colorcoded} and \ref{fig:logage_results_colorcoded} indicate [M/H] (left panel) and [$\mathrm{\alpha/M}$] (right panel). In general, the older age stars have lower [M/H] and higher $[\alpha/\mathrm{M}]$, and the \texttt{BINGO} ages show the clear correlation with [M/H] and $[\alpha/\mathrm{M}]$, as shown in \citet{Ciuca2021}. The \texttt{SIDRA} ages also follow the trend. However, it is not as strong as the trend seen for the \texttt{BINGO} ages, and the stars with similar \texttt{SIDRA} age have a wider variety of [M/H] and [$\alpha$/M]. This ensures that \texttt{SIDRA} is not solely learning the [$\mathrm{\alpha/M}$]-age or [M/H]-age trends, as these trends are more pronounced for \texttt{BINGO} ages.

%%%%%%%%%%%%%%%%%%%%%%%%%%%%%%%%%%%%%%%%%%%%%%%%
%SHAP results
%%%%%%%%%%%%%%%%%%%%%%%%%%%%%%%%%%%%%%%%%%%%%%%%

\begin{figure}
	\includegraphics[width=\columnwidth  ]{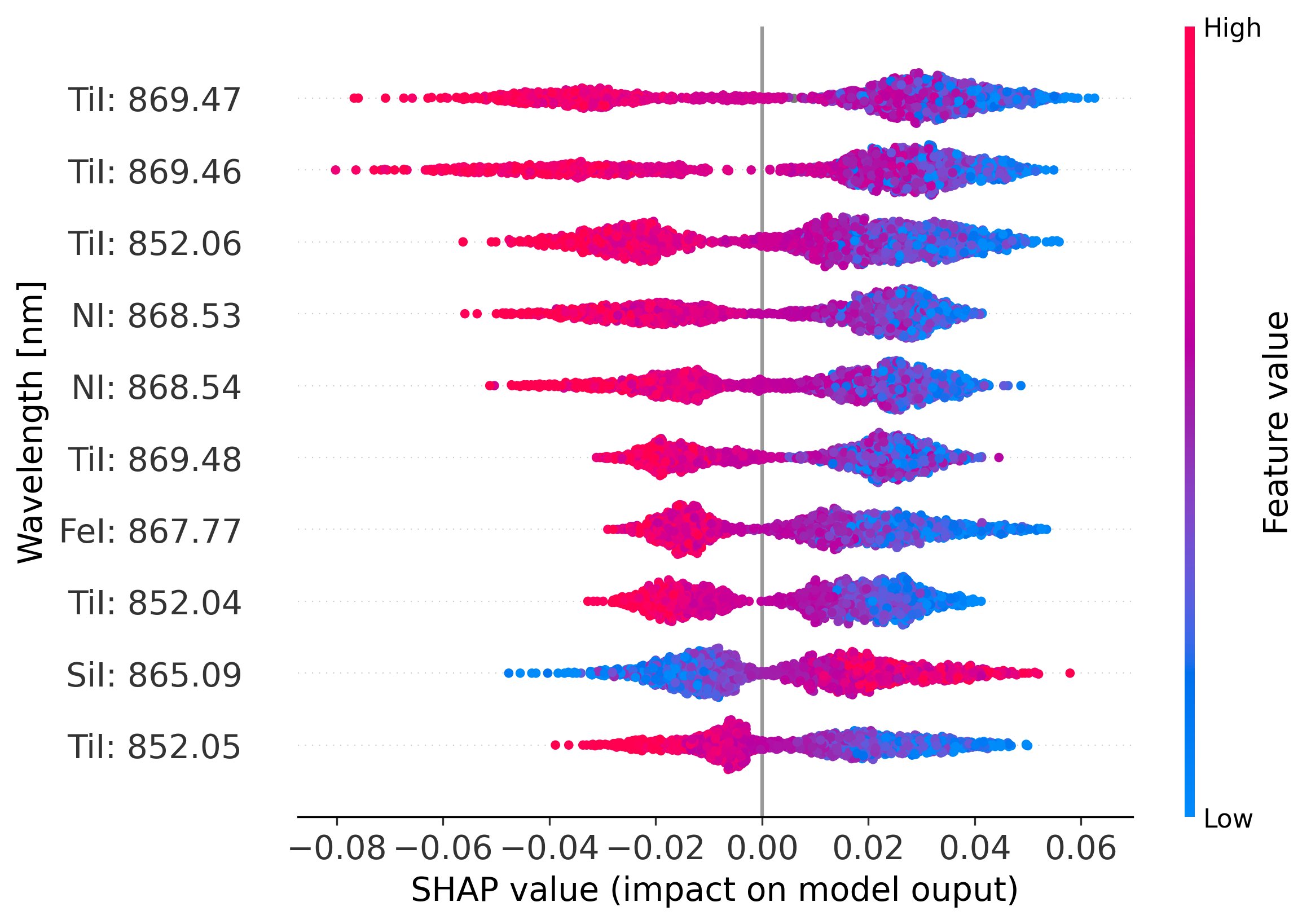}
    \caption{SHAP bee-swarm plot. Each row represents input features at the indicated wavelengths, arranged by significance from top to bottom. Within each row, every point represents a star in the testing dataset, colour-coded by its normalised feature value. The placement of each point illustrates the extent and direction of each feature's influence on its output label, the stellar age, $\mathrm{log}_{10}(\tau_\mathrm{SIDRA}$ $\mathrm{[Gyr]})$.}
    \label{fig:SHAPbeeswarm}
\end{figure}

Fig. \ref{fig:SHAPbeeswarm} is a SHapley Additive exPlanations \citep[SHAP,][]{LundbergLee2017} bee-swarm plot of our model's results shown in Fig. \ref{fig:logage_training_results_colorcoded} and demonstrates the relative importance of the flux values at the different wavelength input features in the RVS spectra and how they impact the stellar-age model output. Fig. \ref{fig:SHAPbeeswarm} shows the distribution of the impact across the top 10 high-impact wavelength pixel values. This visualisation is important for understanding our \texttt{SIDRA-RVS} model's behaviour and the significance of different elemental lines. The x-axis represents the SHAP values, which indicate the impact of each wavelength input feature (normalised flux) on the prediction. SHAP values to the right of the centerline (positive values) suggest that the corresponding input feature contributes to a higher predicted age, whereas values to the left (negative values) indicate that the input feature results in a lower predicted age. The colour of the dots represents the relative values of the input features, with red indicating higher values and blue indicating lower values. Thus, if blue dots appear on the right side, it suggests that lower input feature values correspond to higher output values. The y-axis represents the features sorted by their importance from top to bottom. Here, the importance of the input feature is measured as the summation of the absolute values of SHAP value for all the data. The top elemental features, TiI: 869.47 nm, TiI: 869.46 nm (2nd), and TiI: 869.48 nm (6th), belong to the same absorption line. Similarly, TiI: 852.06 nm (3rd), TiI: 852.05 nm (8th), and TiI: 852.04 nm (10th) are part of the same line. Additionally, NI: 868.53 nm (4th) and NI: 858.54 nm (5th) are also in the same absorption line, making it the third most important line. Blue dots on the high SHAP value side indicate that a lower flux value at the TiI lines, representing a stronger TiI absorption feature in the spectra, predicts a higher age. On the other hand, red dots on the low SHAP value side suggest that a higher flux value at TiI, representing a weaker TiI absorption feature in the spectra, predicts a lower age.

\begin{figure*}
	\includegraphics[width=\textwidth]{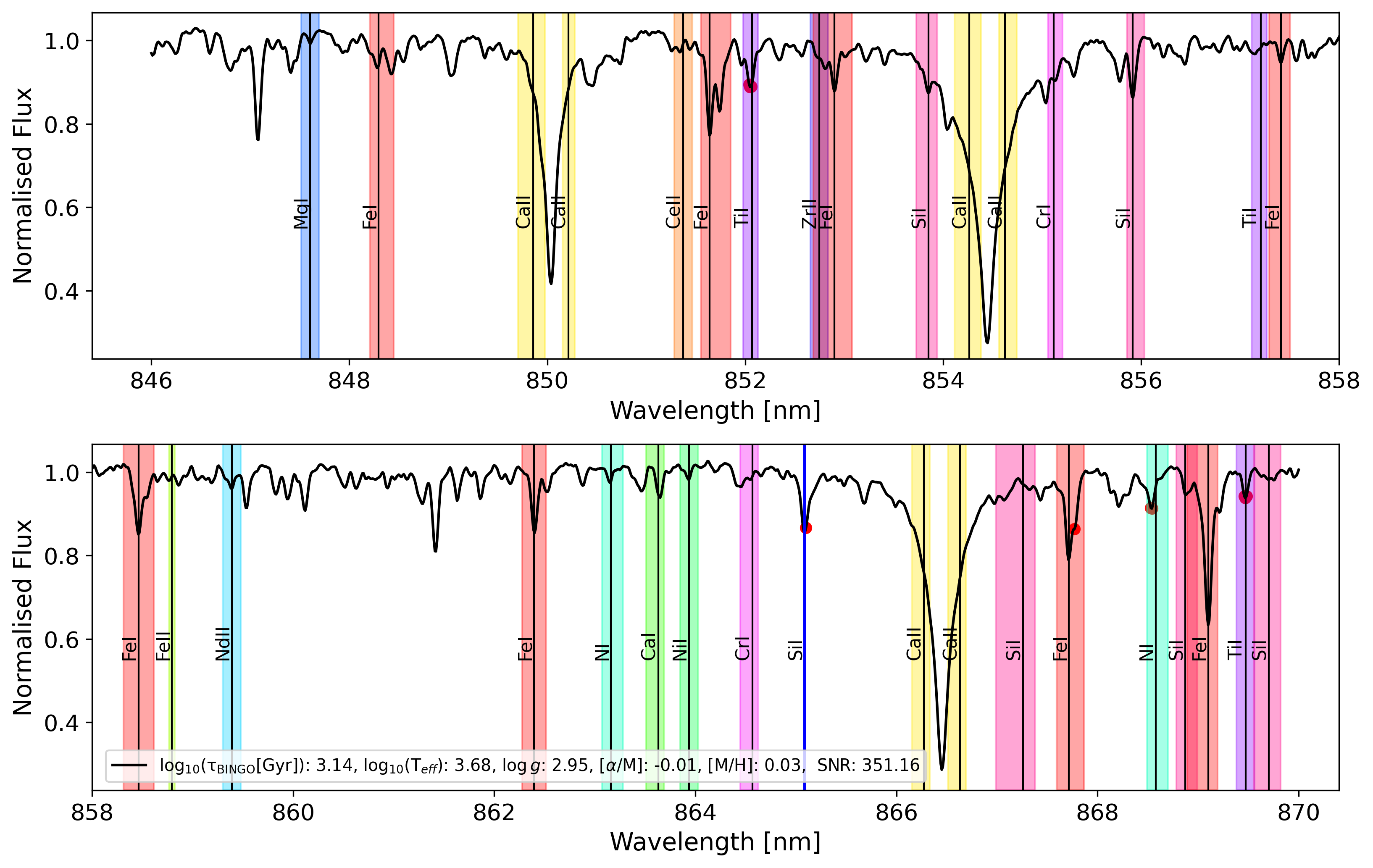}
    \caption{Example of a high signal-to-noise ratio (SNR) spectrum of an observed star from \textit{Gaia} DR3 (DR3 Star ID-1222988540219279360) shown in two wavelength regions: 846-858 nm (top panel) and 858-870 nm (bottom panel). The red dots on the spectrum represent the top 10 highest SHAP values from our \texttt{SIDRA-RVS} model results, overlaid with atomic lines from \citet{RecioBlanco2023}. The width of the atomic lines corresponds to its uncertainty. The vertical blue lines at 865.09 in the bottom panel indicate SiI line identified by \citet{Contursi2021}. The stellar parameters, $\log_{10}{(T_\mathrm{eff})}$, $\log~g$, $[\mathrm{\alpha/M}]$ are obtained from APOGEE.}
    \label{fig:spectra_abun}
\end{figure*}

Fig. \ref{fig:spectra_abun} presents a high signal-to-noise ratio (SNR) spectrum of an observed star from \textit{Gaia}'s RVS as an example. The dots on the spectrum represent the top 10 highest SHAP value features in Fig. \ref{fig:SHAPbeeswarm} from the \texttt{SIDRA-RVS} model results, highlighting the most significant features in the spectrum. The atomic lines from \citet{RecioBlanco2023} are overlaid on the spectrum, with their widths corresponding to their abundance measurement window. The vertical blue lines mark an atomic line identified by \citet{Contursi2021} at wavelengths 865.09 nm in the bottom panel, which was not found in \citet{RecioBlanco2023}'s linelists. This line is identified as a SiI line, offering additional reference from the literature.

Interestingly, our SHAP value analysis identifies the NI line as the third  most elemental significant indicator for stellar age. As discussed in Sec. \ref{sec:intro}, [C/N] is recognised as a reliable age indicator. The accurate age predictions by our model may stem from its consideration of the NI line, and consequently, nitrogen abundance, as important. However, the model does not rank carbon related lines, like CN lines, among the high SHAP value features for age prediction. Fig. \ref{fig:SHAPbeeswarm} reveals that a weaker NI line (higher flux value) results in a lower predicted age, which contradicts the known [C/N]-age relationship where higher nitrogen abundance, i.e. stronger N line, implies higher stellar mass and therefore, a younger age. Hence, it is unlikely that \texttt{SIDRA-RVS} is using nitrogen abundance as an age indicator. On the other hand, Fig. \ref{fig:HRdiag} shows that for the giants focused in this paper, the age is well correlated with the position in the Kiel diagram, and younger stars are higher $T_\mathrm{eff}$ and lower $\log~g$. We therefore think that \texttt{SIDRA-RVS} is learning the information of log \textit{g}, $T_\mathrm{eff}$ and [M/H] from these atomic lines, and the relationship between the stellar age and these stellar parameters. Thus, our model demonstrates that with a carefully selected sample of RGB stars and by limiting the sample of our high-mass RC stars, the RVS spectrum has the power to effectively indicate the ages of these stars.

In Appendix~\ref{sec:appendixB}, we demonstrate that we can train a neural network model with the stellar parameters derived from the RVS spectra in \textit{Gaia}~DR3 \citep[GSP-Spec,][]{RecioBlanco2023}. However, the performance of the trained model purely with the \texttt{GSP-Spec} stellar parameters is worse than \texttt{SIDRA-RVS} which uses the full spectral information.

%%%%%%%%%%%%%%%%%%%%%%%%%%%%%%%%%%%%%%%%%%%%%%%%%%%%%%%%%%%%%%%%%%%%%%%%
\section{SIDRA for the XP stellar parameters: \texttt{SIDRA-XP}}
\label{sec:SIDRA-XP}

\citetalias{FallowsSanders2024} provided the precise stellar parameters, including [C/Fe] and [N/Fe] for giant stars. As described in Sec.~\ref{sec:intro}, [C/N] for giant stars are sensitive to the stellar mass, making these abundance data valuable for predicting the ages of giants. In this section we explore if or not a machine learning model can be trained to infer the precise age from the stellar parameters measured from the XP spectra by \citetalias{FallowsSanders2024}.

%%%%%%%%%%%%%%%%%%%%%%%%%%%%%%%%%%%%%%%%%%%%%%%%%%%%%%%%%%%%%%%%%%%%%%%%
\subsection{\texttt{SIDRA-XP}'s methodology}
\label{subsec:XPmethods}
%%%%%%%%%%%%%%%%%%%%%%%%%%%%%%%%%%%%%%%%%%%%%%%%%%%%%%%%%%%%%%%%%%%%%%%

The relationships between stellar ages and stellar parameters, such as effective temperature, $T_\mathrm{eff}$, surface gravity, log $g$, and various abundance ratios, such as [C/Fe], [Fe/H], [N/Fe] and [$\mathrm{\alpha/M}$], can be complex and non-linear. ANNs are well-suited for this task because they can autonomously identify and model complex relationships without the need for explicitly programmed rules. ANNs are computational models inspired by the human brain's neural networks. They consist of layers of interconnected nodes, or "neurons", which process input data to learn patterns and relationships. In supervised learning, the network's parameters, such as the weights connecting the neurons, are trained and optimised to accurately reproduce known input-output pairs from a training set. Once trained, the ANN can efficiently predict the ages of the other stars based on their observed properties, using significantly less computational effort than the training process.

We split the data 80\% training data and 20\% testing data. To mitigate the data imbalance between young and old stars, we apply the same random-sampling data-augmentation technique described in Sec. \ref{subsec:RVSmethods} to address the underrepresented ages for the training set of the \texttt{BINGO}-FS24. For the data augmentation, we sample $T_\mathrm{eff}$, log $g$, [C/Fe], [Fe/H], [N/Fe], [$\alpha$/Fe] and age from the original star data using the random Gaussian based on the uncertainties in \citetalias{FallowsSanders2024} and \texttt{BINGO}. 

We train an ANN model and optimise the model with $\mathtt{Optuna}$ \citep{Akiba2019+Optuna}. We call our trained model with the XP stellar parameters in \citetalias{FallowsSanders2024} \texttt{SIDRA-XP}. The network architecture includes an input layer that matches the feature size of the training data, two hidden layers with 256 units each, utilising the rectified linear unit (ReLU) activation function to capture non-linear relationships, and an output layer with a single unit and linear activation. The model adopts the the Root Mean Square Propagation ($\mathtt{RMSprop}$) \citep[for more information see,][]{Tieleman2012lecture}, which adjusts the learning rate by averaging squared gradients to improve performance in environments with noisy, sparse, or fluctuating gradients.

%%%%%%%%%%%%%%%%%%%%%%%%%%%%%%%%%%%%%%%%%%%%%%%%%%%%%%%%%%%%%%%%%%%%%%%%
\subsection{Age inference from XP stellar parameters using \texttt{SIDRA-XP} }
\label{subsec:XPresults}
%%%%%%%%%%%%%%%%%%%%%%%%%%%%%%%%%%%%%%%%%%%%%%%%%%%%%%%%%%%%%%%%%%%%%%%%
%%%%%%%%%%%%%%%%%%%%%%%%%%%%%%%%%%%%%%%%%%%%%%%%%%%%%%%%%%%%%%%%%%%%
%Test results colour-coded by [alpha/M] and [M/H] for XP Parameters
%%%%%%%%%%%%%%%%%%%%%%%%%%%%%%%%%%%%%%%%%%%%%%%%%%%%%%%%%%%%%%%%%%%%

\begin{figure*}
    \centering
   \begin{subfigure}[b]{0.495\textwidth}
       \includegraphics[width=\linewidth]{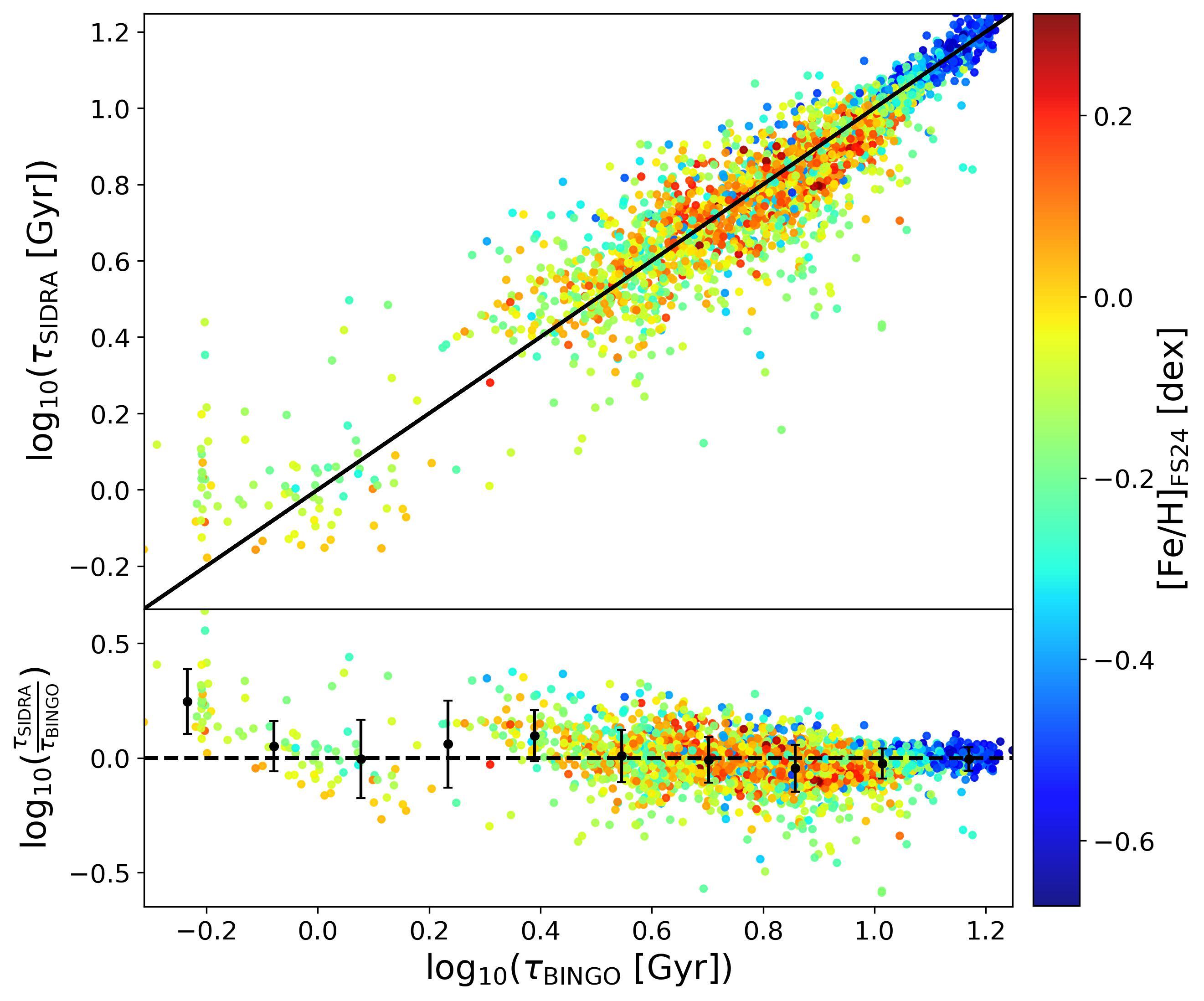}
        \caption{Testing set }
        \label{fig:ANN_test_results_logage_colored_MH}
    \end{subfigure}
    \hfill
    \begin{subfigure}[b]{0.495\textwidth}
        \includegraphics[width=\linewidth]{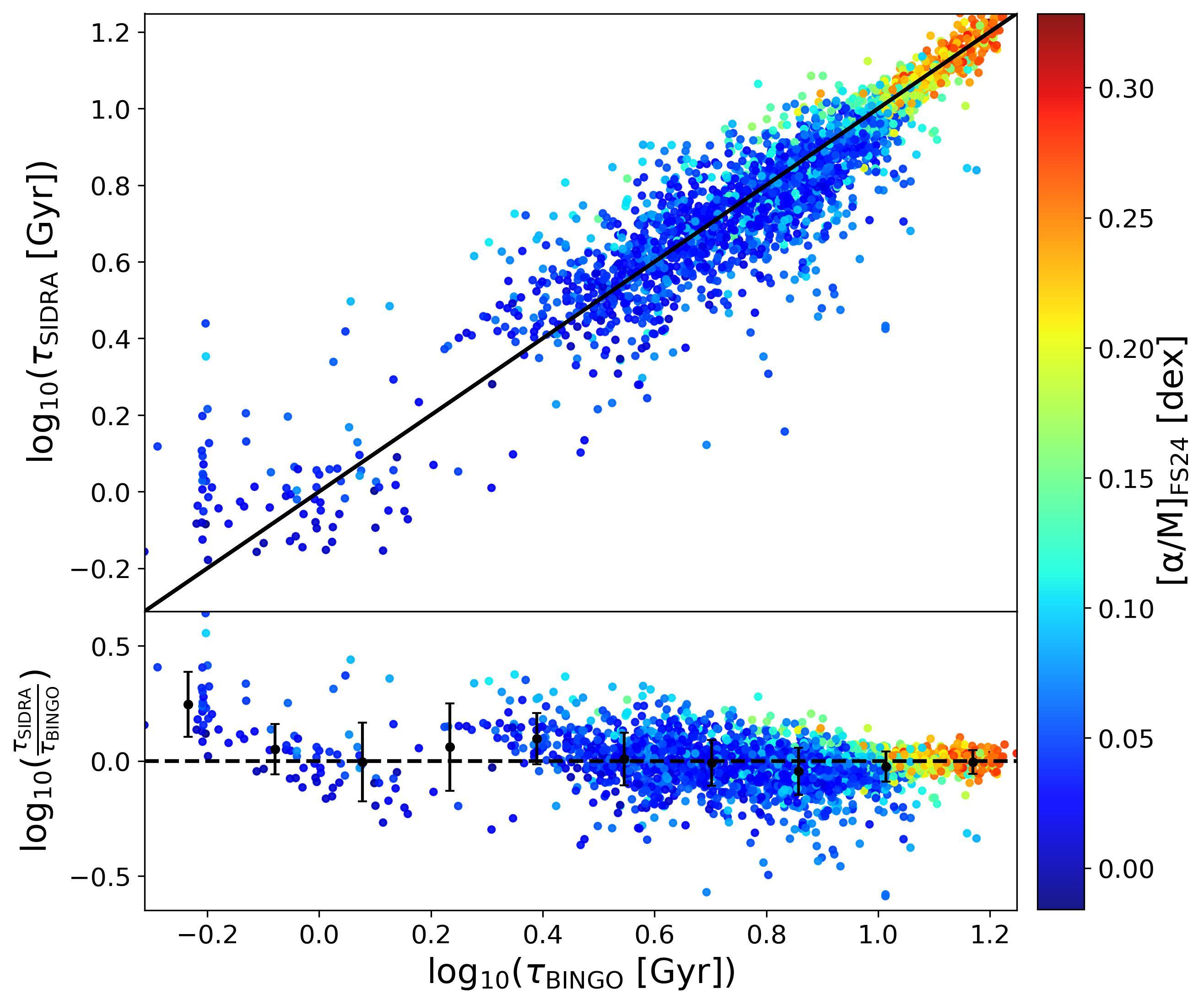}
        \caption{Testing set}
        \label{fig:ANN_test_results_logage_colored_alpham}
    \end{subfigure}
    \caption{\texttt{SIDRA-XP} age predictions, $\mathrm{log}_{10}(\tau_\mathrm{SIDRA}$ $\mathrm{[Gyr]})$, versus the target \texttt{BINGO} age estimation in $\mathrm{log}_{10}(\tau_\mathrm{BINGO}$ $\mathrm{[Gyr]})$ for the XP parameters testing set colour-coded by [Fe/H] (left panel) and [$\mathrm{\alpha/M}$] (right panel) obtained from \citetalias{FallowsSanders2024}. The upper panels show the predictions versus the target and the black line indicates the identity line. The lower panels represent the residuals between the \texttt{SIDRA-XP}'s $\mathrm{log}_{10}(\tau_\mathrm{SIDRA}$ $\mathrm{[Gyr]})$ prediction and \texttt{BINGO}'s true $\mathrm{log}_{10}(\tau_\mathrm{BINGO}$ $\mathrm{[Gyr]})$ denoted as $\mathrm{log}_{10}(\frac{\tau_\mathrm{SIDRA}} {\tau_\mathrm{BINGO}})$. The black filled circles and vertical error bars indicate the mean and the standard deviation of the residuals at the different $\mathrm{log}_{10}(\tau_\mathrm{BINGO}$ $\mathrm{[Gyr]})$ bins, respectively.}
    \label{fig:ANN_logage_results_colorcoded}
\end{figure*}

Fig. \ref{fig:ANN_logage_results_colorcoded} illustrates the performance of the \texttt{SIDRA-XP} trained with the XP stellar parameters provided by \citetalias{FallowsSanders2024} as described in Sec. \ref{sec:data}.  We have split the data to $80\%$ training set and $20\%$ testing set. Fig. \ref{fig:ANN_logage_results_colorcoded} presents the comparison between the ground-truth \texttt{BINGO} ages, $\mathrm{log}_{10}(\tau _{\mathrm{BINGO}}$ $\mathrm{[Gyr]})$, and the corresponding \texttt{SIDRA-XP} predicted ages, $\mathrm{log}_{10}(\tau _{\mathrm{SIDRA}}$ $\mathrm{[Gyr]})$, for the unseen test data. In the testing sets, despite some variations in the predictions, the majority of data points closely follow the identity line. This agreement indicates that \texttt{SIDRA-XP} effectively predicts the ages of the selected giant stars using only the XP stellar parameters from \citetalias{FallowsSanders2024}. 

The bottom panel of Fig. \ref{fig:ANN_logage_results_colorcoded} represents the residuals between \texttt{SIDRA-XP}'s $\mathrm{log}_{10}(\tau _{\mathrm{SIDRA}}$ $\mathrm{[Gyr]})$ prediction and  \texttt{BINGO}'s $\mathrm{log}_{10}(\tau _{\mathrm{BINGO}}$ $\mathrm{[Gyr]})$ age estimates denoted as, $\mathrm{log}_{10}(\frac{\tau_\mathrm{SIDRA}}{\tau_\mathrm{BINGO}})$. Similar to Fig. \ref{fig:logage_results_colorcoded}, the dispersion of residuals is higher for stars with lower age ($\mathrm{log}_{10}(\tau _{\mathrm{BINGO}}$ $\mathrm{[Gyr]})<0.2$) because of the Poisson error due to the smaller number of the young stars in the test data. The mean of the standard deviation of residual is 0.105 dex for stars ($\mathrm{log}_{10}(\tau _{\mathrm{BINGO}}$ $\mathrm{[Gyr]})>0.2$). The standard deviation of the residuals is around 0.064 dex for the test data around $\mathrm{log}_{10}(\tau_\mathrm{BINGO}$ $\mathrm{[Gyr]})= 1$, which is better than the prediction for the \texttt{SIDRA-RVS} testing set as shown in Fig. \ref{fig:logage_results_colorcoded} and discussed in Sec. \ref{subsec:RVSresults}. Our predictions show a larger dispersion compared to the original \texttt{BINGO} data, but interestingly smaller than the RVS data. This clearly demonstrates that the XP parameters themselves have significant power to infer the ages of our selected giants, better than the RVS spectra data.

The colours of the dots in Fig. \ref{fig:ANN_logage_results_colorcoded} indicate [Fe/H] (left panel) and [$\mathrm{\alpha/M}$] (right panel). The figure shows that there is scatter in [Fe/H] and [$\alpha$/M] for the stars with similar age. This means that the \texttt{SIDRA-XP} model does not exclusively capture the correlations between the [$\mathrm{\alpha/M}$] or [Fe/H] and stellar age when using XP parameters, ensuring that the model's learning process goes beyond these specific trends, capturing a broader range of stellar characteristics.

%%%%%%%%%%%%%%%%%%%%%%%%%%%%%%%%%%%%%%%%%%%%%%%%
%SHAP results
%%%%%%%%%%%%%%%%%%%%%%%%%%%%%%%%%%%%%%%%%%%%%%%%

\begin{figure}
	\includegraphics[width=\columnwidth  ]{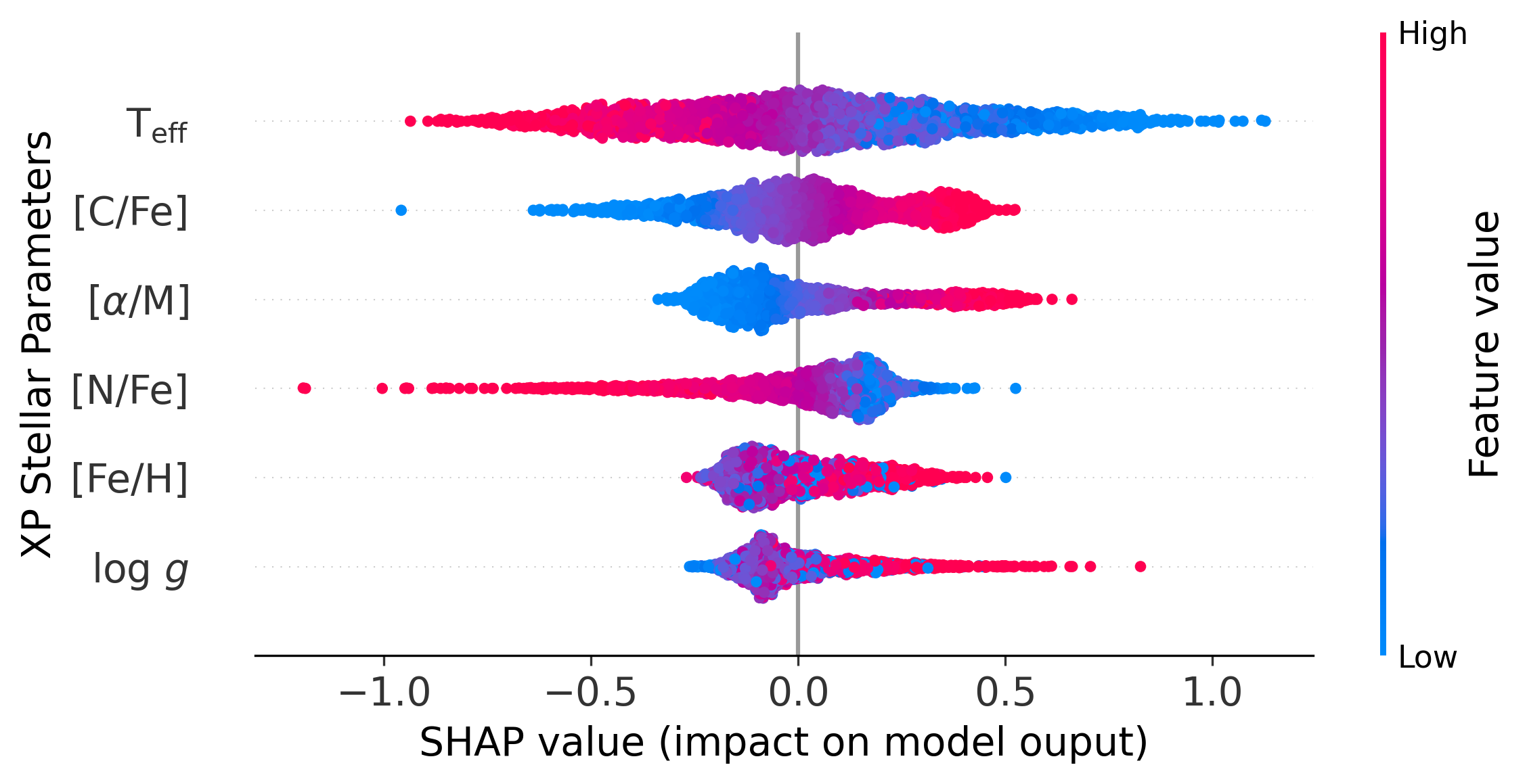}
    \caption{SHAP bee-swarm plot. Each row represents input XP features, arranged by significance from top to bottom. Within each row, every point represents a star in the testing dataset, colour-coded by its normalised feature value. The placement of each point illustrates the extent and direction of each feature's influence on its output label, the stellar age, $\mathrm{log}_{10}(\tau_\mathrm{SIDRA}$ $\mathrm{[Gyr]})$.}
    \label{fig:SHAPbeeswarm_XP}
\end{figure}

Fig. \ref{fig:SHAPbeeswarm_XP} is a SHAP bee-swarm plot that illustrates the contribution of various input features to the predicted ages of stars using \texttt{SIDRA-XP}. Each dot represents an individual star in the test set result as shown in Fig. \ref{fig:ANN_logage_results_colorcoded}. The x-axis indicates the SHAP value, representing the impact of each feature on the model's age prediction for a given star. The features are listed on the y-axis in order from the highest to lowest importance of the input features, which includes effective temperature, $T_\mathrm{eff}$, surface gravity, log $g$, and different abundance ratios, such as [C/Fe], [Fe/H], [N/Fe] and [$\mathrm{\alpha/M}$]. The distribution and spread of the dots indicate how strongly each XP stellar parameter feature influences the stellar age prediction, with dots further from zero having a more significant impact. The alignment of the dots along the x-axis shows whether the features have a positive or negative contribution to the predicted age. In the case of $T_\mathrm{eff}$, the bluer dots in the high SHAP values for $T_\mathrm{eff}$ indicate that low $T_\mathrm{eff}$ stars push the age prediction higher and the redder dots in the lower SHAP values mean that high $T_\mathrm{eff}$ stars push the age prediction to lower. This trend is consistent with the right panel of Fig. \ref{fig:HRdiag} and what is found for \texttt{SIDRA-RVS} in Section~\ref{subsec:RVSresults}.

It is interesting to see that [C/Fe] and [N/Fe] are showing significant impact. For [C/Fe], higher [C/Fe] indicate higher age predictions and lower [C/Fe], indicate lower age predictions. On the other hand, for [N/Fe], an opposite pattern is observed where negative SHAP values, indicate the lower age that correspond to the higher nitrogen abundance. This is consistent with the known age-[C/N] relationship, which suggests that a higher nitrogen abundance compared to the carbon abundance is seen for a higher stellar mass and, consequently, a younger age, as explained in Sec. \ref{sec:intro}. This means that [C/Fe] and [N/Fe] for our selected giant stars are well measured by \citetalias{FallowsSanders2024} with enough accuracy to capture this trend. 

%The [Fe/H] trend of Fig. \ref{fig:SHAPbeeswarm_XP} is puzzling. It shows the higher [Fe/H] leading to older age. However, there are some stars with blue (low [Fe/H]) in the positive SHAP value (older age). Hence, there seem to be a mixed trend of age-[Fe/H] relation. This likely reflects the flat trend of age-[Fe/H] relation in the thin disc, which is also discussed in the next section. 

The [Fe/H] trend of Fig. \ref{fig:SHAPbeeswarm_XP} is puzzling. It shows the higher [Fe/H] leading to older age. However, there are some stars with blue (low [Fe/H]) in the positive SHAP value (older age). Hence, there seems to be a mixed trend of age-[Fe/H] relation. This likely reflects the positive trend of age-[Fe/H] relation in thin disc stars whose age is less than about 8 Gyr, which is also discussed in the next section (see Fig. \ref{fig:3x4density}). On the other hand, the [$\alpha$/M]  shows a clear trend of higher [$\alpha$/M]  leading to older age. This is consistent with the known age-[$\alpha$/M] relation.

Our results demonstrate that the stellar parameters and [C/N] abundances derived from the XP spectra by \citetalias{FallowsSanders2024} provide reliable stellar age information for the giant stars. Unlike \texttt{SIDRA-RVS}, \texttt{SIDRA-XP} learned age-[C/N] relation in addition to the age dependence on $T_\mathrm{eff}$,  $\log~g$ and [$\alpha$/M]. Also, \texttt{SIDRA-XP} shows better performance compared to \texttt{SIDRA-RVS}. Hence, we conclude that the XP stellar parameters measured by \citetalias{FallowsSanders2024} provide the better age estimate than the RVS alone. We also explored using \citetalias{FallowsSanders2024} XP stellar parameters and RVS spectra together with both \texttt{XGBoost} and ANN models. However, we could not see any significant increase in the performance. Because the XP spectra are available for many more stars than the RVS spectra, we conclude that the XP spectra alone would provide the valuable age information for more giant stars. 

Additionally, we explore the case of training \texttt{SIDRA-XP} model without the [C/Fe] and [N/Fe] information. The model still performs effectively, albeit with slightly greater uncertainty compared to when the [C/Fe] and [N/Fe] parameters were included. The standard deviation of the residuals for the test data around $\mathrm{log}_{10}(\tau_\mathrm{BINGO}$ $\mathrm{[Gyr]})= 1$ is approximately 0.068 dex. Consequently, there is approximately 6\% improvement when incorporating the [C/N] data. We compare the \texttt{SIDRA-XP} ages with the stellar ages from \citet{Kordopatis2023} for the same stars. We find that the ages from \citet{Kordopatis2023} do not display the expected correlation with [Fe/H] or [$\alpha$/M], which is typically seen with age as shown in the next section. Consequently, the ages from \citet{Kordopatis2023} appear to be less reliable than the \texttt{SIDRA-XP} ages. However, we stress that this success is likely due to the carefully selected sample of the giant stars in our study following \citet{Ciuca2021}. Our \texttt{SIDRA-XP} is only applicable for the carefully selected populations of the high-mass RC and RGB stars. In the next section, we demonstrate the power of the age estimates of the large number of the giant stars from the \textit{Gaia} XP spectra.

%%%%%%%%%%%%%%%%%%%%%%%%%%%%%%%%%%%%%%%%%%%%%%%%%%%%%%%%%%%%%%%%%%%%%%%%
\section{Chronological map of the Galactic disc with \texttt{SIDRA-XP} }
\label{sec:chronochemicalmap}
%%%%%%%%%%%%%%%%%%%%%%%%%%%%%%%%%%%%%%%%%%%%%%%%%%%%%%%%%%%%%%%%%%%%%%%%

As detailed in Appendix~\ref{sec:appendixC}, we use the trained model, \texttt{SIDRA-XP}, described in Sec. \ref{subsec:XPmethods} for analysing the remainder of the \citetalias{FallowsSanders2024} dataset. As detailed in Appendix~\ref{sec:appendixC}, this yields performance consistent with the model trained on the \citet{Miglio2021} labels. 
%while benefiting from broader coverage in parameter space.
From the \citetalias{FallowsSanders2024} dataset, we select stars that meet the following criteria: $T_\mathrm{eff_{unc}}<69.1$ K, log $g_{\mathrm{unc}}<0.14$ dex, $[\mathrm{Fe}/\mathrm{H}]_{\mathrm{unc}}<0.068$ dex, $4000 < T_{\mathrm{eff}} < 5400$ K and $1.5 < \log g < 3.5$. To exclude low-mass RC stars, we further exclude the stars whose $\log (T_\mathrm{eff} [\mathrm{K}])>3.675$ and $\log g<$2.6~dex.  In cases where duplicate entries exist, the data with the smaller $T_\mathrm{eff_{unc}}$ is selected. 

We further apply the astrometry quality cut for the classified stars, selecting the stars with Renormalised Unit Weight Error, $\texttt{RUWE}<1.4$ \citep{Lindegren2021b} and low parallax error $\mathtt{parallax\_over\_error}>5$. We end up with a total of 2,218,154 stars to use for our \texttt{SIDRA-XP} model.

For our data analysis, the distances to the stars are obtained simply by inverse of the \textit{Gaia}'s parallax measurements \citep{Lindegren2021a}. We also use the zero-point correction of the parallax as suggested by \citet{Lindegren2021b}. We then assume that the Sun's height with respect to the midplane is $z_{\odot}=0.0208$ kpc \citep{Bennett2019} and assume $R_{0}= 8.275$ kpc \citep{GravityCollab2021}. 

\begin{figure*}
	\includegraphics[width=\textwidth]{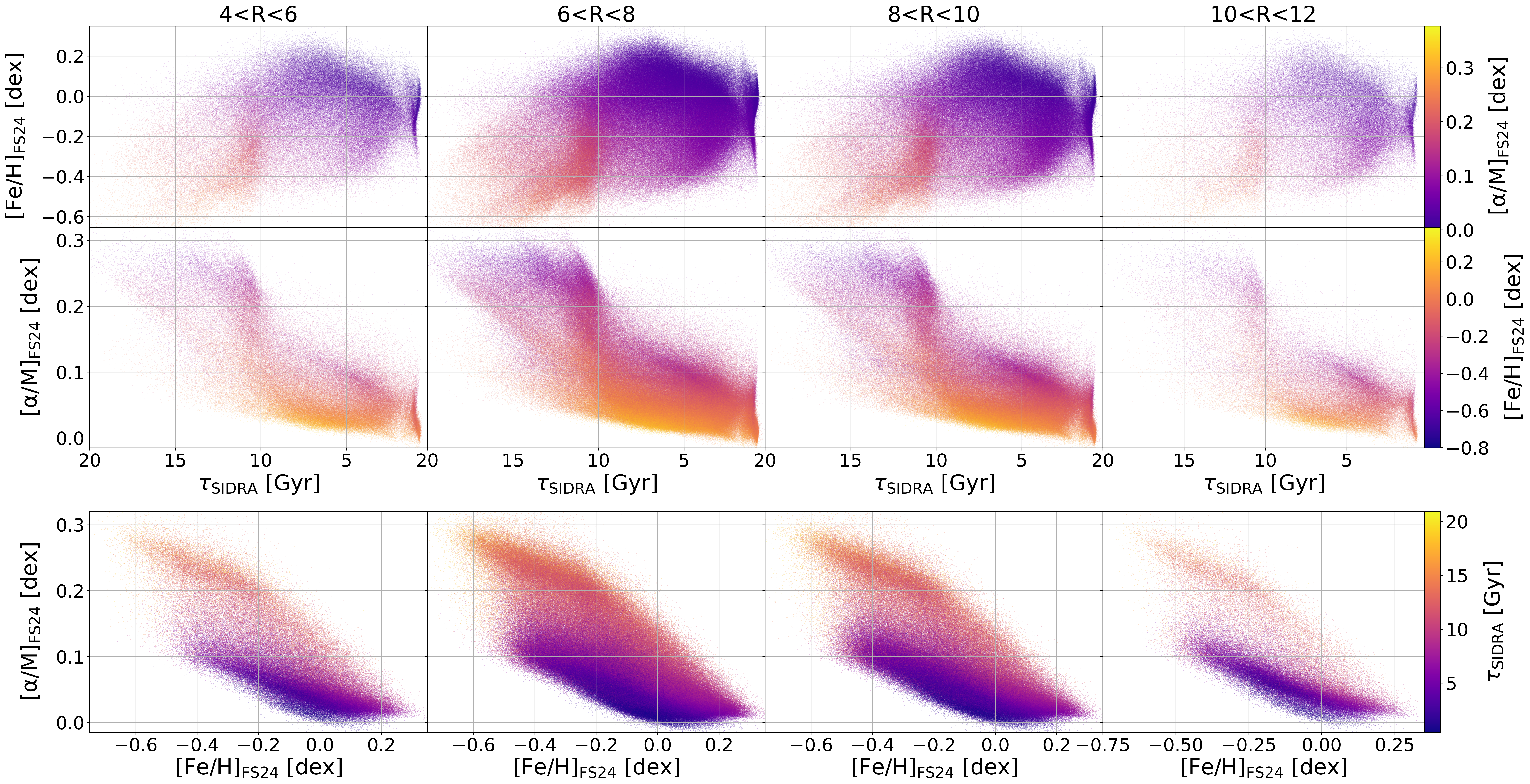}
    \caption{$[\mathrm{Fe/H}]$ as a function of stellar age, $\tau_\mathrm{SIDRA}$, coloured by $[\mathrm{\alpha/M}]$ (top panel), $[\mathrm{\alpha/M}]$ as a function of stellar age, $\tau_\mathrm{SIDRA}$, coloured by metallicity, $[\mathrm{Fe/H}]$ (middle panel) and the distribution of $[\mathrm{\alpha/M}]$ and $[\mathrm{Fe/H}]$, coloured by age, $\tau_\mathrm{SIDRA}$ (bottom panel) across the radial extent of the Galactic disc.}
    \label{fig:3x4}
\end{figure*}

In this section, we examine the evolution of metallicity, $[\mathrm{Fe/H}]$, and $\alpha$-abundances, $[\mathrm{\alpha/M}]$, for the 2,218,154 stars, which were obtained from \citetalias{FallowsSanders2024} XP stellar data as a function of age, derived using \texttt{SIDRA-XP}, $\tau_\mathrm{SIDRA}$. Fig. \ref{fig:3x4} illustrates  $[\mathrm{Fe/H}]$ as a function of $\tau_\mathrm{SIDRA}$, with colours representing $[\mathrm{\alpha/M}]$ (top panel), $[\mathrm{\alpha/M}]$ as a function of $\tau_\mathrm{SIDRA}$, coloured by metallicity, $[\mathrm{Fe/H}]$, (middle panel) and the distribution of $[\mathrm{\alpha/M}]$ and [Fe/H], coloured by age (bottom panel) across the radial extent of the Galactic disc. In the top and middle panels of Fig. \ref{fig:3x4}, the observed deficiency of stars around $\tau_\mathrm{SIDRA} =2.5$ Gyr is due to the selection of high-mass RC stars and a significantly lower number of RGB stars younger than $\sim3$ Gyr. 

The top panel examines the age-metallicity relationship coloured with $[\mathrm{\alpha/M}]$ for \texttt{SIDRA-XP}'s age results. The dataset across the radial extent of the Galactic disc show a flat age-[Fe/H] relation up to approximately 11 Gyr for the younger, low-$[\mathrm{\alpha/M}]$ disc stars, and the expected decrease in [Fe/H] for older, high-$[\mathrm{\alpha/M}]$ stars is also observed in \texttt{SIDRA-XP}'s age results. Our findings qualitatively aligns with the trend seen in the APOGEE data with astroseismic ages \citep [see also][]{SilvaAguirre2018, Mackereth2019, Miglio2021}.

As outlined by \citet{Ciuca2024} and also seen in \citet{Gallart2024, FernandezAlvar2024}, the Galactic disc evolution can be divided into three phases highlighted in the upper panels of Fig. \ref{fig:3x4}. The first phase, which \citet{Ciuca2024} calls Babi, represents an old, metal-poor disc population older than $\tau_\mathrm{SIDRA}>12$ Gyr with a metallicity [M/H]$\simeq-0.4$. The second phase, known as the Great Galactic Starburst (GGS) by \citet{Ciuca2024}, occurs between $9<\tau_\mathrm{SIDRA}<12$ Gyr and is characterised by a rapid increase in metallicity, likely due to a gas-rich merger event of the Gaia-Sausage Enceladus \citep[GSE][]{Belokurov2018, Helmi2018Nature}. During this phase, older stars show lower [Fe/H] and higher $[\mathrm{\alpha/M}]$ ratios, transitioning to higher [Fe/H] and lower $[\mathrm{\alpha/M}]$ as the thin disc begins to form. The third phase is the thin disc formation, continuing from 9 Gyr ago, where younger stars exhibit higher metallicities and lower $[\mathrm{\alpha/M}]$. 

\begin{figure*}
	\includegraphics[width=\textwidth]{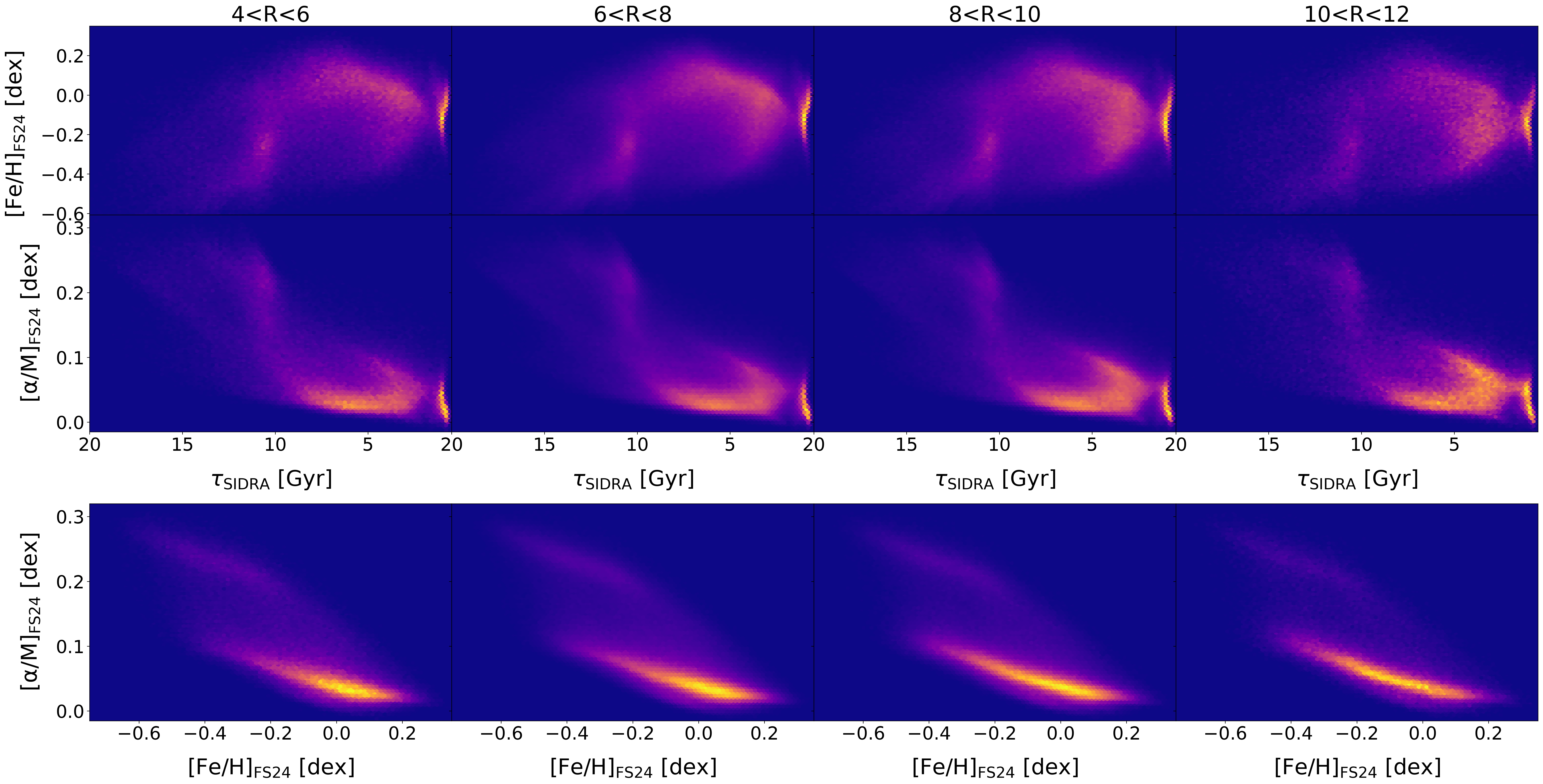}
    \caption{The density plot of $[\mathrm{Fe/H}]$ as a function of stellar age, $\tau_\mathrm{SIDRA}$,  (top panel), $[\mathrm{\alpha/M}]$ as a function of stellar age, $\tau_\mathrm{SIDRA}$, (middle panel), and the distribution of $[\mathrm{\alpha/M}]$ and [Fe/H] (bottom panel) across the radial extent of the Galactic disc .}
    \label{fig:3x4density}
\end{figure*}

The GGS phase is more visible in the inner disc, shown more clearly in the density plot of Fig. \ref{fig:3x4density}, suggesting that the GSE merger which likely involved radial orbits, primarily impacted the inner disc region \citep{Belokurov2018, Helmi2018Nature, Ciuca2024}. This scenario is further supported by \textsc{auriga}, the cosmological simulation work of \citet{Grand2020}, who found that a GSE-like merger played a crucial role in the formation of the Milky Way's thick disc. The merger not only heated existing proto-disc stars, ejecting some into the halo to form the red-main-sequence halo stars referred to as the Splash \citep{DiMatteo2019, Gallart2019, Belokurov2020}, but also brought in fresh gas that triggered a starburst, creating a younger thick disc. The thin disc subsequently began to form post-merger, developing from the gradual accretion of the hot halo gas in an inside-out fashion \citep[i.e.][]{Brook2004,Bird2013,Grand2018, Renaud2021}.

The density plot of Fig. \ref{fig:3x4density} also reveals a noticeable feature around $\tau_\mathrm{SIDRA}$ $\sim$ 7 Gyr with [Fe/H]$\simeq-0.4$. Interestingly, this population also has a higher $[\mathrm{\alpha/M}]$ and creating a sequence of low-[Fe/H] from [Fe/H]$\simeq-0.4$ at $\sim 7$ Gyr to [Fe/H]$\simeq-0.1$ at $\sim$ 2 Gyr in the $\tau_\mathrm{SIDRA}$-[Fe/H] plot, $[\mathrm{\alpha/M}]$ $\simeq0.1$ at $\sim$ 7 Gyr to $[\mathrm{\alpha/M}]$ $\simeq0.05$  at $\sim$ 2 Gyr in the $\tau_\mathrm{SIDRA}$-$[\mathrm{\alpha/M}]$ plot in Figs. \ref{fig:3x4} and \ref{fig:3x4density}. The creation of new stars with low-[Fe/H] and high-$[\mathrm{\alpha/M}]$ ratios is a sign of a gas-rich merger, as it leads to the dilution of \textbf{[Fe/H]} and promotes star formation followed by enrichment from Type II supernovae, which increases $[\mathrm{\alpha/M}]$ as demonstrated in \citet{Brook2007}. These observations suggest that approximately 7 Gyr ago, a satellite galaxy with a significant amount of gas, had a close encounter with the Milky Way gas disc, which triggered a burst of star formation, and introduced metal-poorer gas that later formed new stars. Interestingly, this low-[M/H] population around $\tau_\mathrm{SIDRA}$ $\sim$ 7 Gyr is more pronounced in the outer disc, as also indicated in \citet{Das2020}. Also, this time coincides with the expected first infall time of the Sagittarius dwarf galaxy \citep{Ruiz-Lara2020}. This could be the indication that the Sagittarius galaxy was large enough and the first interaction of the Sagittarius was a gas-rich encounter with the outer edge of the disc. Alternatively, this low metallicity sequence with increasing [Fe/H] for the younger stars could be merely the chemical evolution history at that radius as suggested in \citet{Zhang2025}. The starting point of the metal poor population at age $\sim$ 7~Gyr may indicate that the star formation at the outer disc started after the thin disc becomes large enough during their inside-out formation \citep[e.g.][]{Funakoshi2025}.

The middle panel of Fig. \ref{fig:3x4} illustrates the age-$[\mathrm{\alpha/M}]$ relationship coloured with metallicity for \texttt{SIDRA-XP}'s age results.  The high-$[\mathrm{\alpha/M}]$ ‘cluster’ separates clearly from the low-$[\mathrm{\alpha/M}]$ ‘cluster’ in the age-$[\mathrm{\alpha/M}]$ space at $[\mathrm{\alpha/M}]$ $\sim0.15$ dex. Most high-$[\mathrm{\alpha/M}]$ stars ($[\mathrm{\alpha/M}]$ > $0.15$ dex) are generally older and more metal-poor ($[\mathrm{Fe/H}]< -0.2$ dex) compared to the relatively younger low-$[\mathrm{\alpha/M}]$ population which is shown across all Galactic radial ranges.

The bottom panel of Fig. \ref{fig:3x4} shows the distribution of [Fe/H] and $[\mathrm{\alpha/M}]$ of our disc stars coloured by age, $\tau_\mathrm{SIDRA}$, across different radial ranges of the Galactic disc. In this panel, there is a clear trend where younger stars tend to have higher [Fe/H], while older stars are more metal-poor. The \texttt{SIDRA-XP} ages also display a distinction between high-$[\mathrm{\alpha/M}]$ and low-$[\mathrm{\alpha/M}]$ stars across different ages. Older stars tend to have higher $[\mathrm{\alpha/M}]$ abundances, while younger stars show lower $[\mathrm{\alpha/M}]$ abundances. Overall, \texttt{SIDRA-XP} captures the overall age trend from high-$[\mathrm{\alpha/M}]$ and low-[Fe/H] thick disc populations to low-$[\mathrm{\alpha/M}]$ and high-[Fe/H] thin disc population.

For low-[$\alpha$/Fe] and high-[Fe/H] thin disc population generally shows a broad but flat [Fe/H] distribution in the age-[Fe/H] relation (upper panels in Figs.~ \ref{fig:3x4} and \ref{fig:3x4density}). Interestingly, in addition to the negative age-[Fe/H] sequence for lower [Fe/H] stars as discussed above, there is a positive age-[Fe/H] sequence for higher [Fe/H] from [Fe/H]$\sim$0.2 at $\tau_\mathrm{SIDRA}\sim9$~Gyr to [Fe/H]$\sim$0.1 at $\tau_\mathrm{SIDRA}\sim2$~Gyr as visible in Fig.~\ref{fig:3x4density}. This positive trend of the age-[Fe/H] relation is likely a driver of the higher SHAP value (older age) for the higher [Fe/H] feature in Fig. \ref{fig:SHAPbeeswarm_XP} and the lower SHAP value (younger age) for the lower [Fe/H] feature.  This trend is also seen in previous literature. For example, \citet{XiangandRix2022} interpret this feature as a consequence of radial migration, where fewer young, metal-rich stars born in the inner disc migrate to the outer disc. However, we observe this negative age-[Fe/H] sequence in the wide range of the radius. This may require an alternative scenario to explain. Similar conclusions were also drawn by \citet{Nepal2024}. 

The abundance of XP data in \citetalias{FallowsSanders2024} also allows a high quality of age predictions to reproduce the known age trends. Observing these trends across a vast number of bright giant stars indicates that, with careful consideration of the selection function, \texttt{SIDRA-XP}'s age predictions for an even larger sample of giant stars in future \textit{Gaia} data releases will significantly advance our understanding of the Milky Way's structure and evolution.

%trend is very much aligned with Hayden et al and Ciuca et al. paper It is good to mention what is known in APOGEE data and Anders paper 

%\section{Conclusions}
\section{Summary}
\label{sec:summary}

This paper demonstrates that it is feasible to deduce stellar ages from \textit{Gaia}'s RVS spectra using the \texttt{XGBoost} algorithm, we call \texttt{SIDRA-RVS}, particularly for RGB and high-mass RC stars by training the model with the carefully selected APOGEE giant data in \citet{Ciuca2024}. Our results indicate that the \texttt{SIDRA-RVS} model predicts stellar ages with high accuracy, achieving a standard deviation of residuals of approximately 0.12 dex for a test set of stars with ages around 10 Gyr. The model's performance is robust, indicating that \textit{Gaia}'s RVS spectra can effectively estimate stellar ages. Based on the SHAP bee-swarm analysis, we find that some of TiI, NI  and FeI lines are significant indicators of age, with lower flux values (stronger absorption features) generally predicting higher ages. Interestingly, while nitrogen is a significant predictor, none of the high SHAP value features correspond to CN lines. We suspect that the model learned the correlations between the age and the stellar parameters of $T_\mathrm{eff}$, $\log$ \textit{g} and [M/H] from these lines. 

In addition to our analysis with RVS spectra, we trained \texttt{SIDRA} using stellar parameters derived from \textit{Gaia}'s XP spectra provided by \citetalias{FallowsSanders2024}, denoted as \texttt{SIDRA-XP}. This model leverages stellar parameters, such as $T_\mathrm{eff}$, $\log$ \textit{g} and chemical abundance ratios, to infer stellar ages. Our findingd demonstrate that \texttt{SIDRA-XP} achieves even better precision, with a standard deviation of residuals around 0.064 dex for the unseen test data with ages around 10 Gyr, outperforming \texttt{SIDRA-RVS}. The SHAP analysis of \texttt{SIDRA-XP} confirms that $T_\mathrm{eff}$, $\log$ \textit{g} and abundance ratios, especially [C/Fe] and [N/Fe], are key predictors of stellar age. This suggests that \texttt{SIDRA-XP} effectively captures the well-established correlations between age and stellar parameters, including the [C/N] ratio, which has been shown to correlate with stellar mass and hence age.

To illustrate the utility of \texttt{SIDRA-XP}, we apply the trained model to a large sample of 2,218,154 stars from \textit{Gaia}'s XP stellar parameter data, selecting stars that meet specific accuracy criteria from \citet{FallowsSanders2024}. Our analysis mapped the chronological and chemical evolution of the Galactic disc, revealing distinct phases in the disc's history: an early metal-poor phase, a rapid increase in metallicity likely linked to the GSE merger, and the ongoing formation of the thin disc. The results provide a hint of a gas-rich merger event around 7 Gyr ago, potentially associated with the first infall of the Sagittarius dwarf galaxy, as indicated by a sequence of stars with lower [Fe/H] and higher $[\mathrm{\alpha/M}]$.

Overall, this study highlights the potential of \textit{Gaia}'s RVS and XP spectra to provide valuable age information for stars, especially giants, and emphasises the importance of careful sample selection to achieve precise age estimates. By leveraging machine-learning techniques, such as \texttt{SIDRA}, we can unlock the full potential of large spectroscopic data from the future data releases of the \textit{Gaia} mission, enhancing our understanding of the formation and evolution of the Milky Way.

\section*{Acknowledgements}

We would like to thank our anonymous referees for their detailed assessment and insightful recommendations. ASA acknowledges the funding body, the UAE Ministry of Presidential Affairs, for their support through the PhD scholarship. This work was partly supported by the UK's Science \& Technology Facilities Council (STFC grant ST/S000216/1, ST/W001136/1).
JLS acknowledges support from the Royal Society (URF\textbackslash R1\textbackslash191555).
AM acknowledges support from the ERC Consolidator Grant funding scheme (project ASTEROCHRONOMETRY,  G.A. n. 772293).
This work is a part of MWGaiaDN, a Horizon Europe Marie Sk\l{}odowska-Curie Actions Doctoral Network funded under grant agreement no. 101072454 and also funded by UK Research and Innovation (EP/X031756/1). This work has made use of data from the European Space Agency (ESA) mission {\it Gaia} (\url{https://www.cosmos.esa.int/gaia}), processed by the {\it Gaia}
Data Processing and Analysis Consortium (DPAC, \url{https://www.cosmos.esa.int/web/gaia/dpac/consortium}). Funding for the DPAC has been provided by national institutions, in particular the institutions participating in the {\it Gaia} Multilateral Agreement. Funding for the Sloan Digital Sky Survey IV has been provided by the Alfred P. Sloan Foundation, the U.S. Department of Energy Office of Science, and the Participating Institutions. SDSS-IV acknowledges support and resources from the Center for High Performance Computing  at the University of Utah. The SDSS website is \url{www.sdss4.org} SDSS-IV is managed by the Astrophysical Research Consortium for the Participating Institutions of the SDSS Collaboration including the Brazilian Participation Group, the Carnegie Institution for Science, Carnegie Mellon University, Center for Astrophysics | Harvard \& Smithsonian, the Chilean Participation Group, the French Participation Group, Instituto de Astrof\'isica de Canarias, The Johns Hopkins University, Kavli Institute for the Physics and Mathematics of the 
Universe (IPMU) / University of Tokyo, the Korean Participation Group, Lawrence Berkeley National Laboratory, Leibniz Institut f\"ur Astrophysik Potsdam (AIP),  Max-Planck-Institut f\"ur Astronomie (MPIA Heidelberg), Max-Planck-Institut f\"ur Astrophysik (MPA Garching), 
Max-Planck-Institut f\"ur Extraterrestrische Physik (MPE), National Astronomical Observatories of China, New Mexico State University, 
New York University, University of Notre Dame, Observat\'ario Nacional/MCTI, The Ohio State University, Pennsylvania State University, Shanghai Astronomical Observatory, United Kingdom Participation Group, Universidad Nacional Aut\'onoma de M\'exico, University of Arizona, University of Colorado Boulder, University of Oxford, University of Portsmouth, University of Utah, University of Virginia, University of Washington, University of Wisconsin, Vanderbilt University, and Yale University.

%%%%%%%%%%%%%%%%%%%%%%%%%%%%%%%%%%%%%%%%%%%%%%%%%%
\section*{Data Availability}
The datasets used in this article were obtained from publicly available sources: \textit{Gaia}, DR3: \url{https://gea.esac.esa.int/archive/}; \citetalias{FallowsSanders2024}; \url{https://zenodo.org/records/12660210}. 

The output catalogue produced by this work is available at \url{https://zenodo.org/records/14260592}.



\bibliographystyle{mnras}
\bibliography{ms} % if your bibtex file is called example.bib

%\begin{thebibliography}{99}
%\end{thebibliography}

%%%%%%%%%%%%%%%%%%%%%%%%%%%%%%%%%%%%%%%%%%%%%%%%%%

%%%%%%%%%%%%%%%%% APPENDICES %%%%%%%%%%%%%%%%%%%%%

\appendix
\section{\textit{Gaia}~DR3 \texttt{FLAME} ages }
\label{sec:appendixA}

\begin{figure*}
    \centering
   \begin{subfigure}[b]{0.495\textwidth}
        \includegraphics[width=\linewidth]{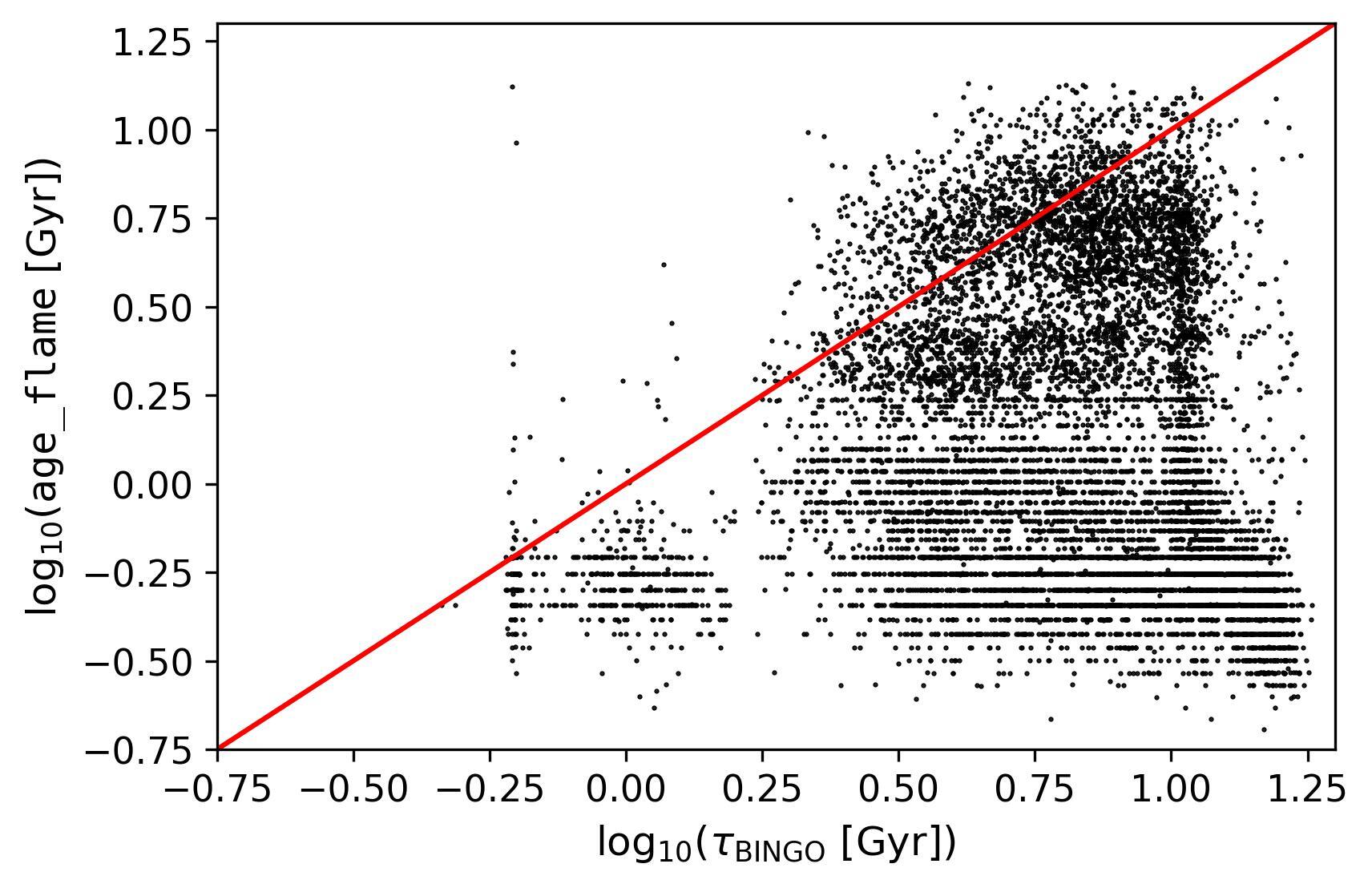}
        \label{subfig:age_flame}
    \end{subfigure}
    \hfill
    \begin{subfigure}[b]{0.495\textwidth}
        \includegraphics[width=\linewidth]{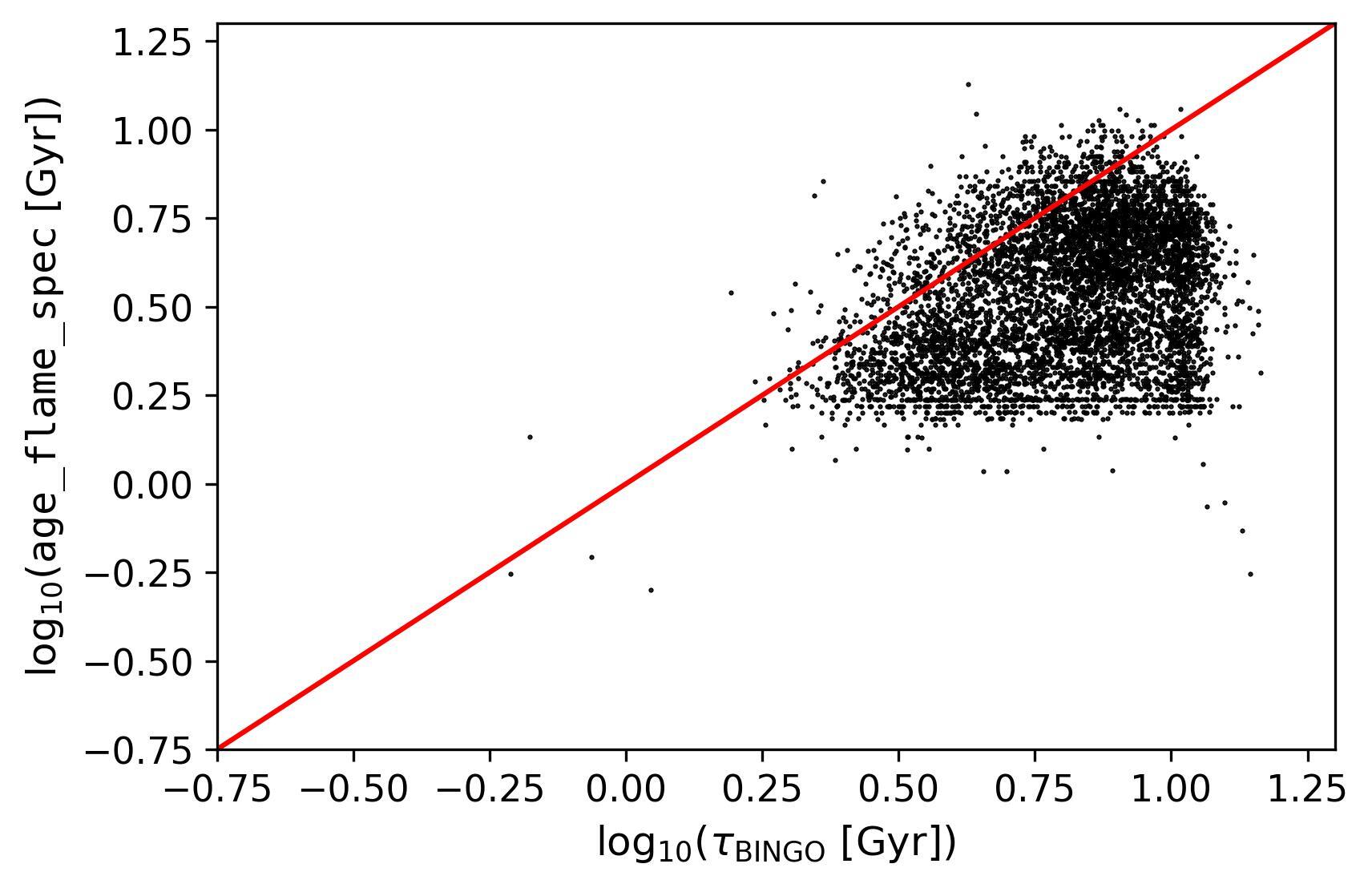}
        \label{subfig:age_flame_spec}
    \end{subfigure}
    \caption{ Comparison between 
    the stellar agess obtained from \texttt{BINGO}, $\mathrm{log}_{10}(\tau_\mathrm{BINGO}$ $\mathrm{[Gyr]})$, and $\mathtt{age\_flame}$, $\log_{10}(\mathtt{age\_flame}\mathrm{[Gyr]})$  (left panel) and $\mathtt{age\_flame\_spec}$, $\log_{10}(\mathtt{age\_flame\_spec} \mathrm{[Gyr]})$ (right panel) data obtained from \textit{Gaia} DR3 for red giant branch and high-mass red clump stars. The red solid line represents the identity line.}
    \label{fig:age_gaia}
\end{figure*}

We cross-match the \texttt{BINGO} data with \textit{Gaia} DR3 data with $\mathtt{age\_flame}$ ages and obtain 10,507 stars. We show their age comparison in the left panel of Fig.~\ref{fig:age_gaia}. We also cross-match the \texttt{BINGO} data with \textit{Gaia} DR3 data with $\mathtt{age\_flame\_spec}$ ages and found 5,957 stars with $\mathtt{age\_flame\_spec}$ ages. The right panel of Fig.~\ref{fig:age_gaia} presents their age comparison. There are 12,500 stars in \texttt{BINGO}-FS24 and 8,859 stars in \texttt{BINGO}-RVS (Sec.~\ref{sec:data}). The \textit{Gaia} \texttt{FLAME} ages are available for fewer stars than in our sample and also show significant underestimates of the stars.

%%%%%%%%%%%%%%%%%%%%%%%%%%%%%%%

\section{Model trained with the Gaia Gsp-Spec stellar parameters}
\label{sec:appendixB}

\begin{figure*}
    \centering
   \begin{subfigure}[b]{0.495\textwidth}
       \includegraphics[width=\linewidth]{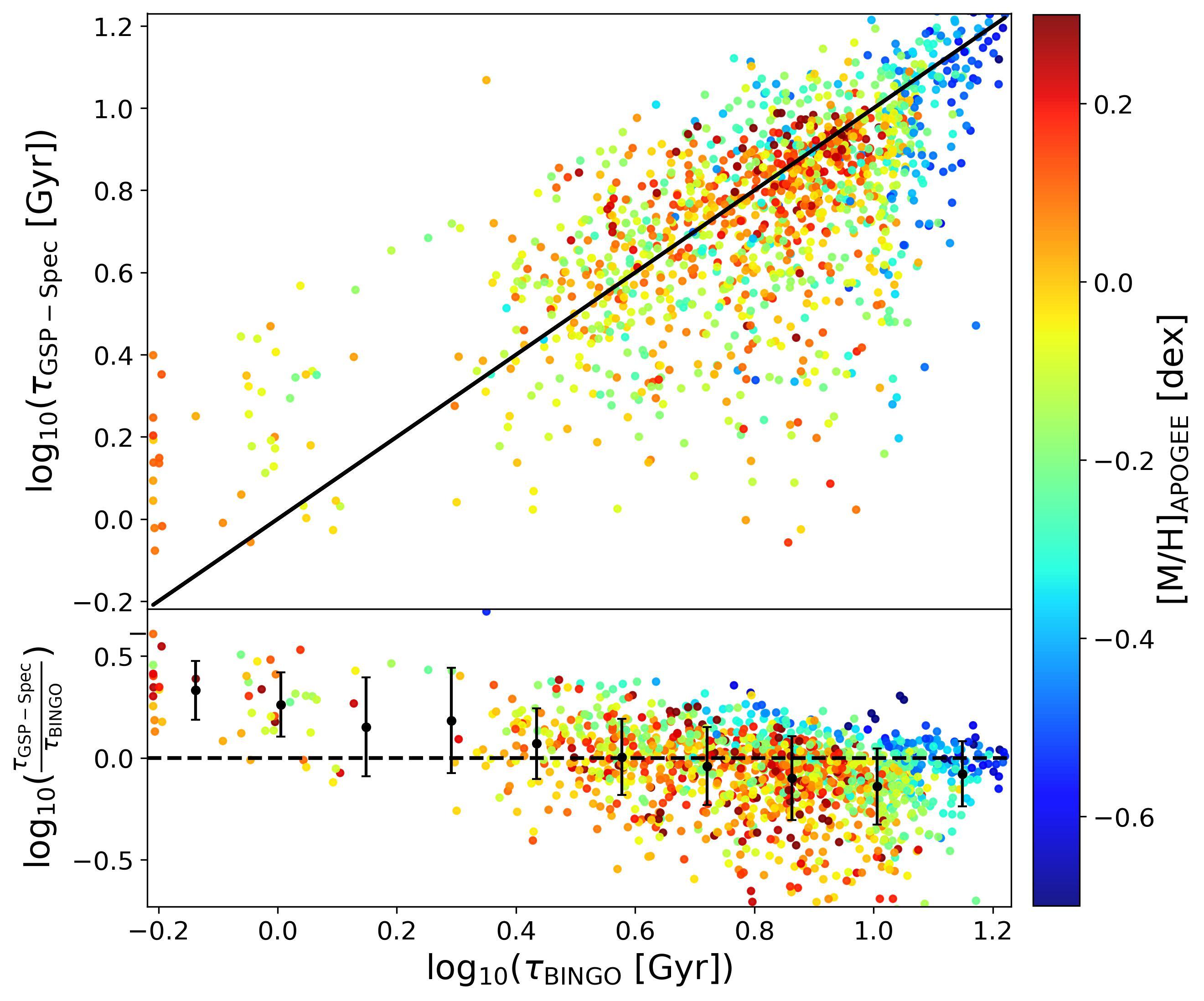}
        % \caption{Model trained with the GSP-Spec data.}
        % \label{fig:ANN_test_results_GspSpec}
    \end{subfigure}
    \hfill
    \begin{subfigure}[b]{0.495\textwidth}
        \includegraphics[width=\linewidth]{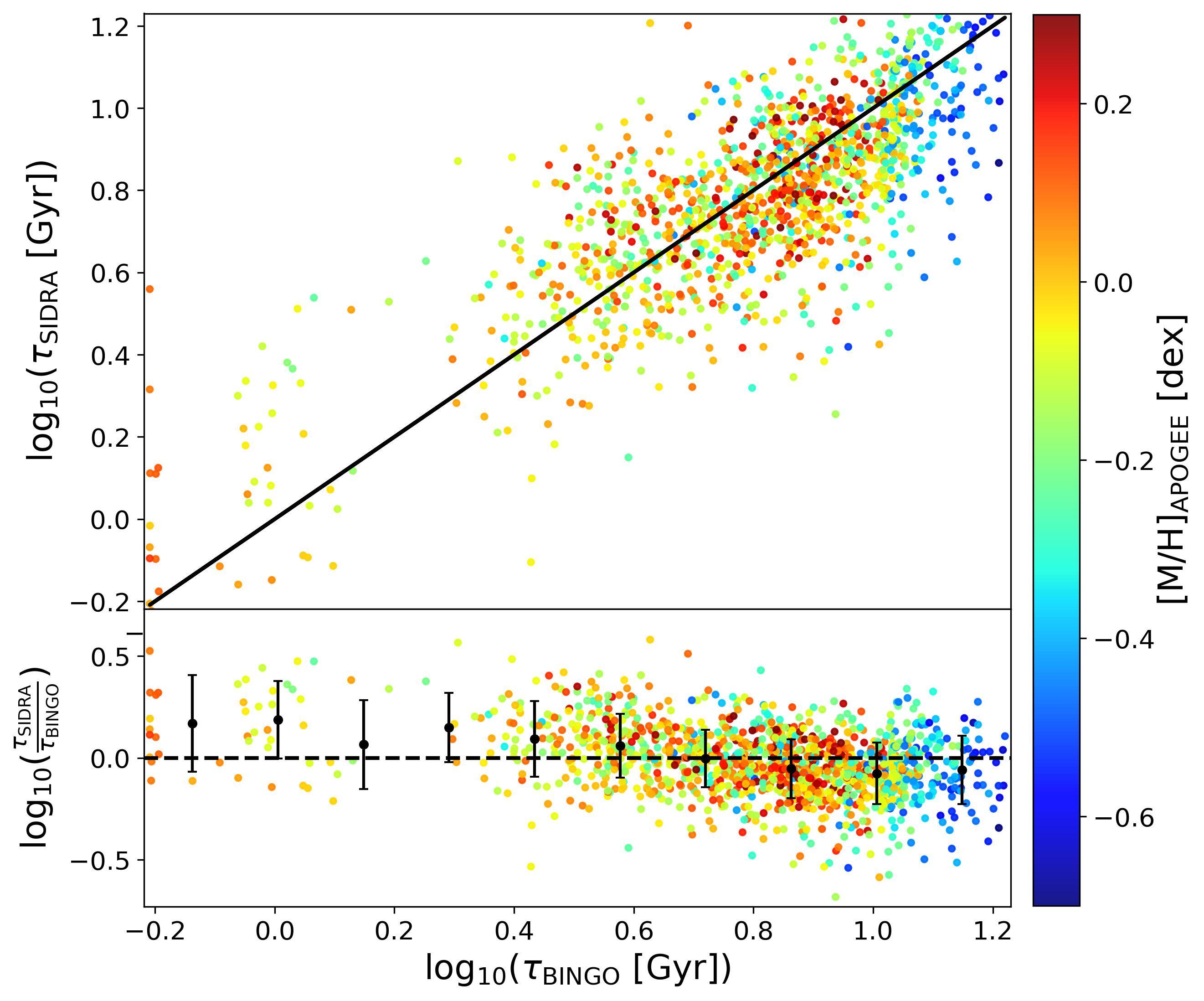}
        % \caption{\texttt{SIDRA-RVS} trained with the RVS spectra of the GSPSpec-\texttt{BINGO} data.}
        %\label{fig:XGBoost_test_results_MH_RVSxGspSpec}
    \end{subfigure}
    \caption{Age predictions, $\mathrm{log}_{10}(\tau_\mathrm{SIDRA}$ $\mathrm{[Gyr]})$, from the GSP-Spec stellar parameters versus the target \texttt{BINGO} age estimation in $\mathrm{log}_{10}(\tau_\mathrm{BINGO}$ $\mathrm{[Gyr]})$ (left panel). The prediction of the age for the same sample as the left panel, but trained with \texttt{SIDRA-RVS} are shown in the right panel. The results are colour-coded by [M/H] parameter obtained from APOGEE. The upper panels show the predictions versus the target and the black line indicates the identity line. The lower panels represent the residuals between the predicted age and the target age. The black filled circles and vertical error bars indicate the mean and the standard deviation of the residuals at different $\mathrm{log}_{10}(\tau_\mathrm{BINGO}$ $\mathrm{[Gyr]})$ bins.}
    \label{fig:GspSpec_Vs_RVS}
\end{figure*}

The \textit{Gaia}~DR3 provides stellar parameters derived from the RVS spectra \citep[GSP-Spec,][]{RecioBlanco2023}. We cross-match the \texttt{BINGO} data with \textit{Gaia}~DR3 stars that have GSP-Spec measurements for the effective temperature ($T_{\mathrm{eff, GSP-Spec}}$), surface gravity ($\log g_{\mathrm{GSP-Spec}}$), and metallicity ($[\mathrm{M/H}]_{\mathrm{GSP-Spec}}$), resulting in a dataset of 6,924 stars, referred to as the \texttt{BINGO}-GSP-Spec dataset. An artificial neural network (ANN) model is trained using 80\% of these cross-matched data, with $T_{\mathrm{eff, GSP-Spec}}$, $\log g_{\mathrm{GSP-Spec}}$ and $[\mathrm{M/H}]_{\mathrm{GSP-Spec}}$ as the input features and \texttt{BINGO} age as the output. We apply the same data augmentation techniques to the training data as described in Sec. \ref{subsec:RVSmethods} and optimise the hyperparameters using \texttt{Optuna}. The predicted age for the unseen test data—i.e., the 20\% of the original dataset not used for training—is shown in the left panel of Fig.~\ref{fig:GspSpec_Vs_RVS}. The ANN model trained on GSP-Spec parameters shows significant residuals, indicating discrepancies between the predicted and true age estimates ($\mathrm{log}_{10}(\tau_\mathrm{SIDRA}$ [Gyr])). This suggests that the model struggles to differentiate between younger and older stars, with the residuals showing considerable spread (mean of approximately 0.20 dex) and uncertainty (standard deviation of $\sim0.18$ dex around $\mathrm{log}_{10}(\tau_\mathrm{BINGO}$ [Gyr]) = 1).

To ensure a fair comparison, we retrain \texttt{SIDRA-RVS}, an \texttt{XGBoost} model trained with RVS spectra data, using 80\% of the \texttt{BINGO}-GSP-Spec dataset. The result is shown in the right panel of Fig.~\ref{fig:GspSpec_Vs_RVS}. While the age recovery in the left panel is reasonable, there is more scatter in the predicted ages compared to the \texttt{SIDRA-RVS} model’s results for the test data. The standard deviation of the residuals for the RVS test data is approximately 0.15 dex around $\log_{10}(\tau_\mathrm{BINGO}$ [Gyr]) = 1, which is better than the corresponding predictions in the left panel. Thus, we conclude that directly using the RVS spectra improves the age prediction.

\section{Performance Assessment of \texttt{SIDRA-XP} Models with APOKASC3 Data}
\label{sec:appendixC}

\begin{figure*}
    \centering
   \begin{subfigure}[b]{0.495\textwidth}
       \includegraphics[width=\linewidth]{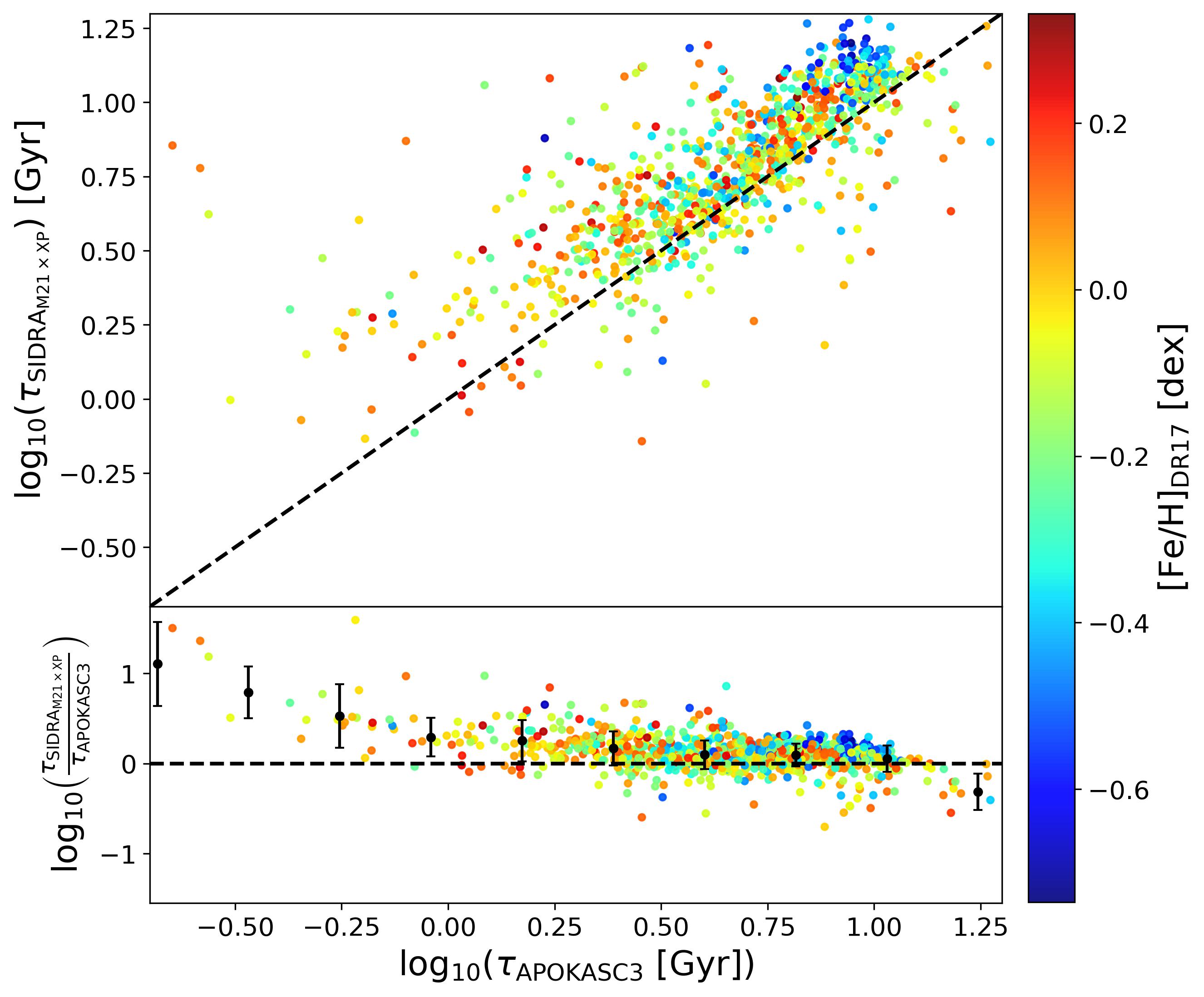}
    \end{subfigure}
    \hfill
    \begin{subfigure}[b]{0.495\textwidth}
        \includegraphics[width=\linewidth]{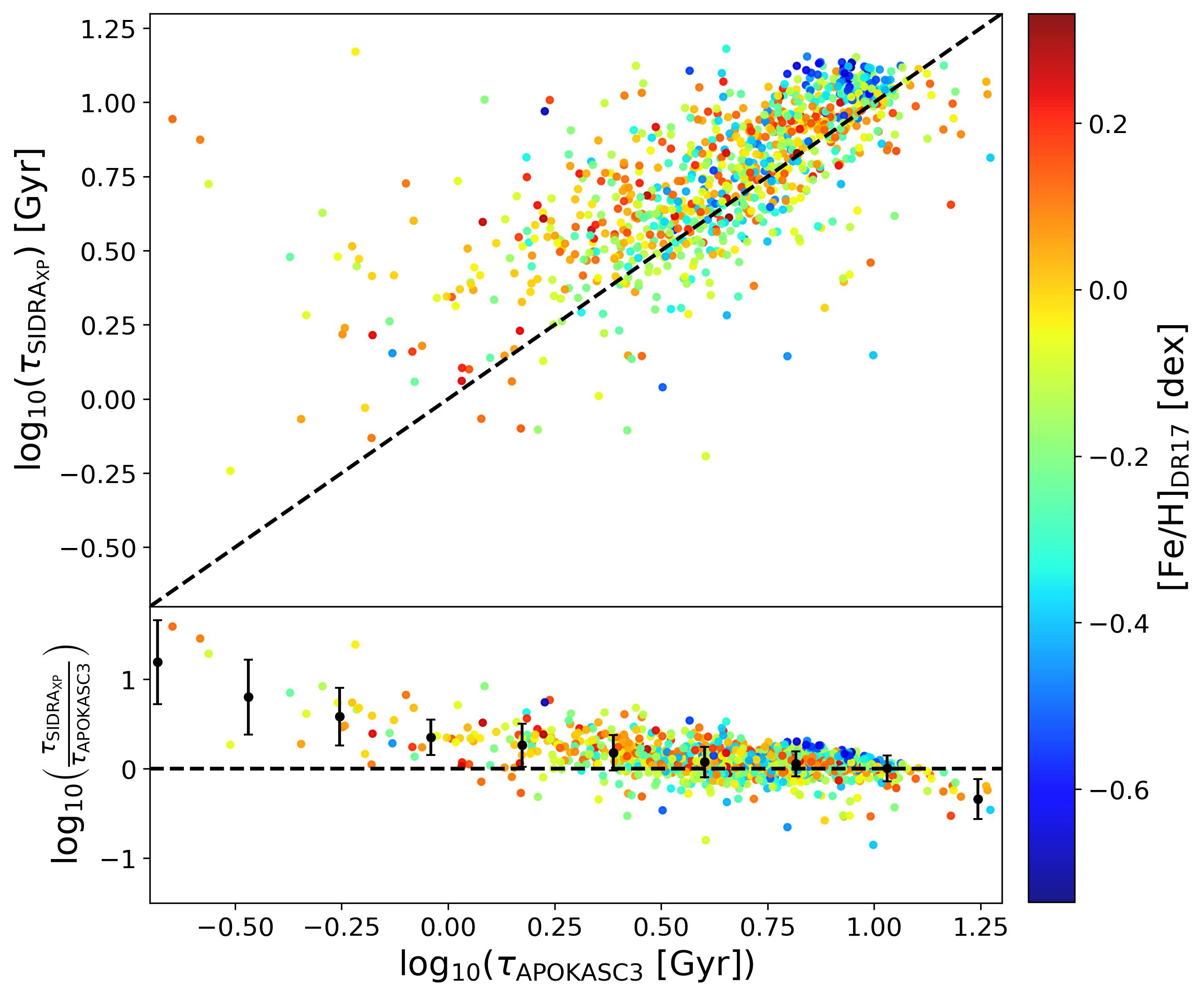}
    \end{subfigure}
    \caption{Age predictions, $\mathrm{log}_{10}(\tau_\mathrm{SIDRA_{M21\times XP}}$ $\mathrm{[Gyr]})$, from \citetalias{FallowsSanders2024} XP stellar parameters trained on age labels from \citet{Miglio2021} versus the target APOKASC3 age estimation in $\mathrm{log}_{10}(\tau_\mathrm{APOKASC3}$ $\mathrm{[Gyr]})$ (left panel). The prediction of the age for the same sample as the left panel, but trained with \texttt{SIDRA-XP} using the \texttt{BINGO} \citet{Ciuca2021} age labels are shown in the right panel. The results are colour-coded by [Fe/H] parameter from APOGEE DR17. The upper panels show the predictions versus the target and the black dashed line indicates the identity line. The lower panels represent the residuals between the predicted age and the target age, i.e. $\log_{10}(\tau_{\mathrm{SIDRA}} / \tau_{\mathrm{APOKASC3}})$. The black filled circles and vertical error bars indicate the mean and the standard deviation of the residuals at different $\mathrm{log}_{10}(\tau_\mathrm{APOKASC3}$ $\mathrm{[Gyr]})$ bins.}
    \label{fig:SIDRAM21xXP}
\end{figure*}

This appendix presents a comparative evaluation of two independently trained versions of the \texttt{SIDRA} model, as the model introduced in Section \ref{sec:SIDRA-XP}. Both models utilise the XP stellar parameters from \textit{Gaia} DR3, as analysed in \citetalias{FallowsSanders2024}, but differ in the training datasets for age calibrations. One model, referred to here as \texttt{SIDRA-M21xXP}, was trained using the asteroseismic ages published by \citet{Miglio2021}, with the training set restricted to include stars identified as RGB stars or RC stars with masses exceeding 1.8 $\mathrm{M}_\odot$. The other model, \texttt{SIDRA-XP}, was trained using the more extensive \texttt{BINGO}  dataset from \citet{Ciuca2021}, which itself is calibrated on the ages from \citet{Miglio2021} as described in Section~\ref{subsec:XPmethods}. 

Although the \citet{Miglio2021} sample is smaller than that of \texttt{BINGO}, this comparison allows us to test the consistency and robustness of \texttt{SIDRA} predictions under different training regimes. We aim to assess the trade-off between a pristine but smaller number of the training data (2695 stars) of \citet{Miglio2021} sample and a larger training set (12,500 stars) produced with \texttt{BINGO}. To ensure consistency, \texttt{SIDRA-M21xXP} model was also constructed following the same methodology outlined in Section~\ref{subsec:XPmethods}.

To evaluate their predictive performance, we apply both models to a subset of the APOGEE–Kepler Asteroseismic Science Consortium Catalog, version 3 \citep[APOKASC3;][]{Pinsonneault2025}, ensuring no overlap with either training set from \texttt{SIDRA-M21xXP} and \texttt{SIDRA-XP}, which leads to 970 stars. The testing sample was constructed by selecting all stars identified as RGB, and for which reliable asteroseismic age estimates were available in APOKASC3. For each selected star, we adopt the \texttt{APOKASC3\_AGE\_RGB}.

\begin{figure}
    \centering
    \includegraphics[width=\columnwidth]{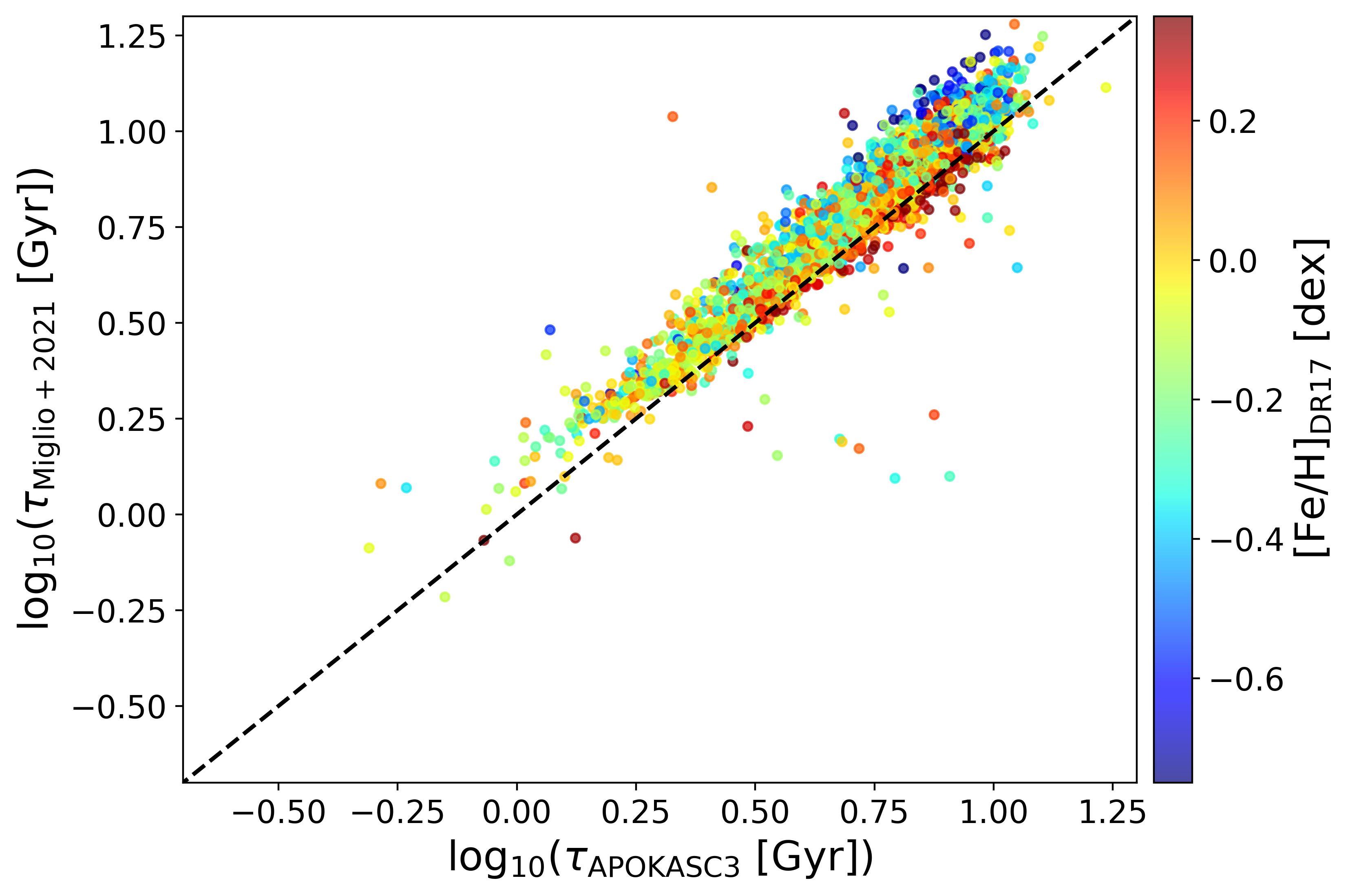}
    \caption{Comparison between $\log_{10}$ ages from \citep[APOKASC3;][]{Pinsonneault2025} and those from \citet{Miglio2021}, for RGB and high-mass RC stars colour-coded by APOGEE [Fe/H]$_\mathrm{DR17}$. The black dashed line indicates the identity line.}
    \label{fig:Miglio_vs_AP3}
\end{figure}

Comparisons between the model-predicted ages and the APOKASC3 values reveal that both models achieve comparable performance as shown in Fig. \ref{fig:SIDRAM21xXP}. Both models infer the systemically older ages than the APOKASC3 ages. This is because the asteroseismic age of \citet{Miglio2021} is systematically older than the APOKASC3 age as shown in Fig. \ref{fig:Miglio_vs_AP3} Interestingly, the \texttt{SIDRA} model trained on \citetalias{FallowsSanders2024} XP and \texttt{BINGO} data exhibited slightly improved accuracy and reduced residual scatter in the overall sample, i.e. the standard deviation of the residuals is around 0.14 dex for $\mathrm{log}_{10}(\tau_\mathrm{APOKASC3}$ $\mathrm{[Gyr]})=1$ for \texttt{SIDRA-XP} and 0.15 for $\mathrm{log}_{10}(\tau_\mathrm{APOKASC3}$ $\mathrm{[Gyr]})=1$ for \texttt{SIDRA-M21xXP}. 
% This is because a larger number of training data with \texttt{BINGO} increases the confidence level of the mean data trend becomes higher. This means that \texttt{SIDRA-XP} model is forcing to follow the main trend of the original training data more strongly. It is still assuring to see that the inferred age of \texttt{SIDRA-XP} are more similar to independently measured APOKASC3 age.  
We consider that this difference is not significant, and rather indicate that both training data provide a similar quality of the trained model. % In this paper, we adopt \texttt{SIDRA-XP} trained with the BINGO sample, applying it to infer stellar ages for giant stars in \textit{Gaia} DR3.

While conducting this analysis, we observed that both models showed discrepancies for stars with APOKASC3 ages below 1 Gyr. We find that these APOKASC3 stars have high [C/N], but with a high mass. Hence, APOKASC3 inferes a younger age, but our model infers an older age.
% Closer inspection of the age distribution [C/N], colour-coded by [Fe/H], revealed that the 
Fig.~\ref{fig:Miglio_vs_AP3} shows that \citet{Miglio2021} training set includes very few RGB stars in this young age regime. Consequently, the model trained on that dataset struggles to recover accurate ages for APOKASC3 young stars,
% , particularly those with higher [C/N] ratios. We also found that these APOKASC3 stars tend to have high inferred masses, 
highlighting the challenge of reliable extrapolation in this regime. Although their ages may be overestimated and may contaminate older populations, their rarity suggests that the overall impact on population-level inferences is likely limited. However, this caveat should be acknowledged and kept in mind when interpreting the results.

\begin{figure}
    \centering
    \includegraphics[width=\columnwidth]{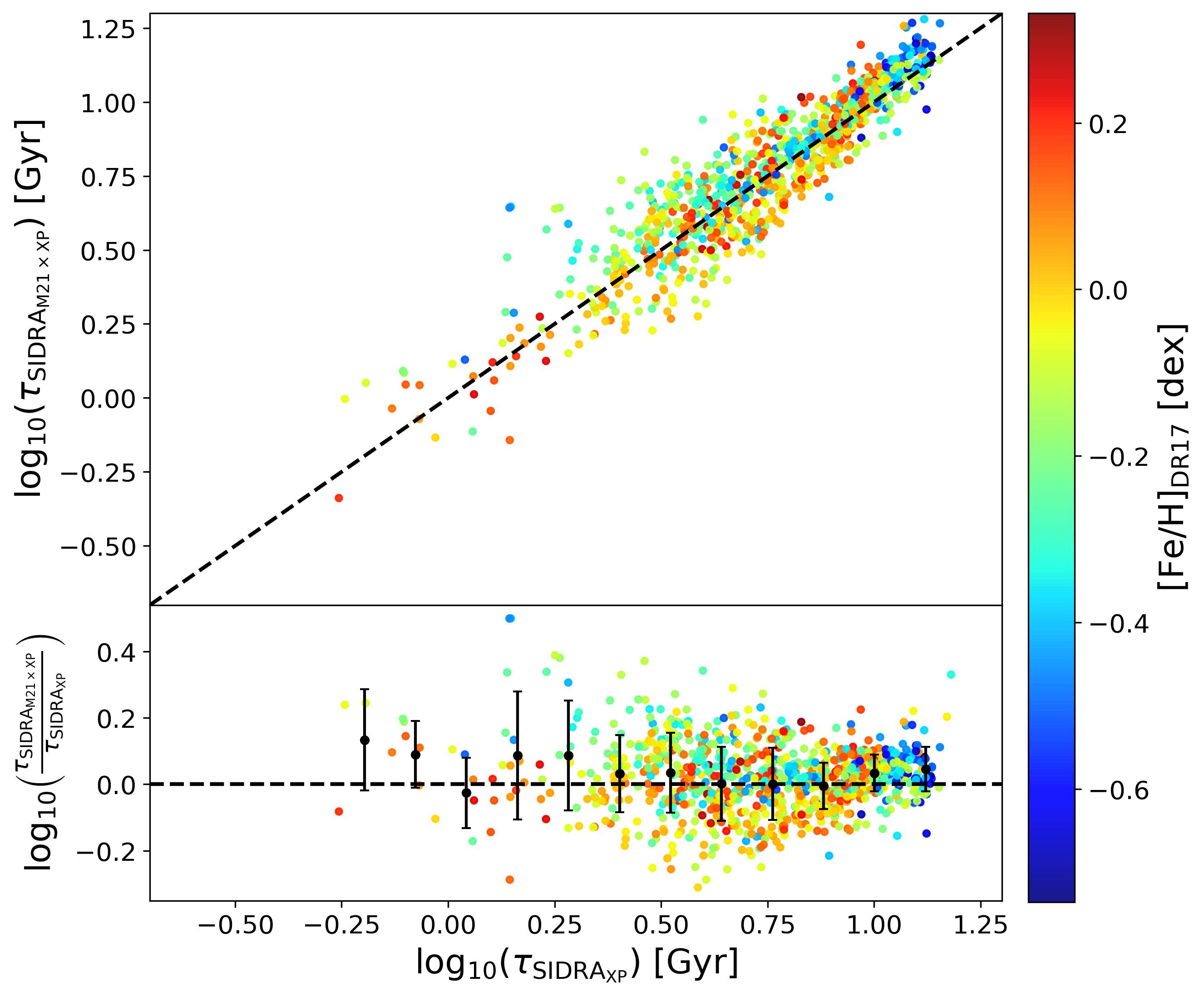}
    \caption{
        Comparison between the stellar ages predicted by two independently trained models: \texttt{SIDRA-M21xXP} and \texttt{SIDRA-XP}. Both models use \textit{Gaia} DR3 XP parameters from \citetalias{FallowsSanders2024} as input, but differ in their age labels: \citet{Miglio2021} for \texttt{SIDRA-M21xXP} and the \texttt{BINGO} dataset from \citet{Ciuca2021} for \texttt{SIDRA-XP}. The results are colour-coded by [Fe/H] parameter from APOGEE DR17. The top panel shows a 1:1 comparison of the models' respective age predictions, $\mathrm{log}_{10}(\tau_\mathrm{SIDRA_{M21\times XP}}\,\mathrm{[Gyr]})$ and $\mathrm{log}_{10}(\tau_\mathrm{SIDRA_{XP}}\,\mathrm{[Gyr]})$, with the black dashed line indicating the identity line. The lower panel shows the residuals between the two models, i.e. $\mathrm{log}_{10}(\tau_\mathrm{SIDRA_{M21\times XP}} / \tau_\mathrm{SIDRA_{XP}})$. The black filled circles and vertical error bars indicate the mean and the standard deviation of the residuals at different $\mathrm{log}_{10}(\tau_\mathrm{SIDRA_{XP}})$ bins.
    }
    \label{fig:SIDRA_M21xXP_VS_SIDRA_XP}
\end{figure}

Fig.~\ref{fig:SIDRA_M21xXP_VS_SIDRA_XP} illustrates the close agreement between the stellar age predictions from the \texttt{SIDRA-M21xXP} and \texttt{SIDRA-XP} models, with residuals tightly centred around zero and consistent performance across the full age and metallicity range. These results indicate that the choice of age calibration, whether using \citet{Miglio2021} or \texttt{BINGO}, does not substantially affect the performance of the model when using XP spectra as input. In this paper, we adopt \texttt{SIDRA-XP} trained on the \texttt{BINGO} dataset, but this plot shows that \texttt{SIDRA-M21xXP} yields comparably reliable predictions, supporting the robustness of the method.

% If you want to present additional material which would interrupt the flow of the main paper,
% it can be placed in an Appendix which appears after the list of references.

%%%%%%%%%%%%%%%%%%%%%%%%%%%%%%%%%%%%%%%%%%%%%%%%%%

% Don't change these lines
\bsp	% typesetting comment
\label{lastpage}
\end{document}